\RequirePackage[hyphens]{url}
\documentclass[preprint,12pt]{elsarticle}

\usepackage{graphicx}
\usepackage{amssymb}
\usepackage{lineno}
\usepackage{color}
\usepackage{wrapfig}
\usepackage{amsmath}
\usepackage{xcolor}
\usepackage{adjustbox}
\usepackage{subcaption}
\usepackage{tabularx}
\usepackage{float}
\usepackage{hyperref}
\usepackage{doi}
\usepackage{listings}
\usepackage{color}
\usepackage{caption}
\definecolor{codegreen}{rgb}{0,0.6,0}
\definecolor{codegray}{rgb}{0.5,0.5,0.5}
\definecolor{codepurple}{rgb}{0.58,0,0.82}
\definecolor{backcolour}{rgb}{0.95,0.95,0.92}

\lstdefinestyle{mystyle}{
    backgroundcolor=\color{backcolour},
    commentstyle=\color{codegreen},
    basicstyle=\ttfamily\scriptsize
}

\DeclareMathOperator*{\argmin}{arg\,min}
\DeclareMathOperator*{\argmax}{arg\,max}
\biboptions{sort&compress}
\biboptions{semicolon}

\usepackage{natbib}
\setcitestyle{authoryear,open={(},close={)}}

\journal{Epidemics}

\begin{document}

\begin{frontmatter}

\title{Contemporary statistical inference for infectious disease models using Stan}

\author[1]{Anastasia Chatzilena\footnote{Corresponding author: Anastasia Chatzilena, xatzilenan@aueb.gr}}
\author[2]{Edwin van Leeuwen}
\author[3]{Oliver Ratmann}
\author[4,5]{Marc Baguelin}
\author[6,7]{Nikolaos Demiris}

\address[1]{Department of Economics, Athens University of Economics and Business, Athens, Greece}
\address[2]{Respiratory Diseases Department, Public Health England, London, United Kingdom}
\address[3]{Department of Mathematics, Imperial College London, United Kingdom}
\address[4]{School of Public Health, Infectious Disease Epidemiology, Imperial College London, United Kingdom}
\address[5]{Centre for Mathematical Modelling of Infectious Disease and Department of Infectious Disease Epidemiology, London School of Hygiene and Tropical Medicine, London, UK}
\address[6]{Department of Statistics, Athens University of Economics and Business, Athens, Greece}
\address[7]{Cambridge Clinical Trials Unit, University of Cambridge, Cambridge, UK}

\begin{abstract}
This paper is concerned with the application of recent statistical advances to inference of infectious disease dynamics. We describe the fitting of a class of epidemic models using Hamiltonian Monte Carlo and Variational Inference as implemented in the freely available Stan software. We apply the two methods to real data from outbreaks as well as routinely collected observations. Our results suggest that both inference methods are computationally feasible in this context, and show a trade-off between statistical efficiency versus computational speed. The latter appears particularly relevant for real-time applications. 
\end{abstract}


\end{frontmatter}

\section{Introduction}
\label{S:1}
The dynamics of infectious diseases depend on how the balance of uninfected and infected individuals varies over time. In all but the most simplest cases mathematical modelling is an indispensable tool for understanding the resulting epidemic spread. However, fitting epidemic models is not straightforward, typically because the actual numbers of uninfected (susceptible) and infected individuals remain unobserved, which we refer to as being latent from a statistical perspective. In this context, Bayesian approaches to modelling and inference of infectious disease dynamics have the advantage that latent parameters and their uncertainties can be seamlessly accounted for. However exploiting this principal advantage is often made difficult by substantial challenges in developing computational tools that work efficiently in a broad range of infectious disease applications. The BUGS software \cite[]{lunn2000winbugs} is one example of such computational tools, automating numerical inference and providing an easy-to-use interface for building and sharing Bayesian statistical models. Other, more recently developed examples of such general purpose tools for computational fitting of Bayesian models are JAGS \cite[]{jags2017}, Nimble \cite[]{de2017programming}, AD Model Builder \cite[]{fournier2012ad}, Template Model Builder \cite[]{kristensen2015tmb} and PyMC \cite[]{patil2010pymc}. Yet, many recent Bayesian modelling approaches in infectious disease epidemiology rely on highly customised Markov Chain Monte Carlo (MCMC) and adaptive MCMC methods for learning the model parameters from data \cite[]{baguelin2013assessing, o1999bayesian}. This state of play is a major hindrance for developing, sharing and fitting mathematical models to characterize the spread of infectious diseases.

As model complexity increases, the performance of classical MCMC algorithms deteriorates due to their potentially inefficient exploration of the target distribution. The latest developments in statistics and machine learning suggest that Hamiltonian Monte Carlo (HMC) methods \cite[]{2017arXiv170102434B,2012arXiv1206.1901N} and Variational Bayes (VB) \cite[]{blei2017variational,kucukelbir2017automatic} may offer increased statistical and/or computational efficiency compared to MCMC. The relatively new software package Stan \cite[]{Stan2018Stan, carpenter2017Stan} provides a generic interface to implementing both HMC and VB, freeing end-users from the challenge of implementing their own computational HMC and VB routines. In addition, it appears that Stan is the first such software offering built-in solvers for systems of ordinary differential equations (ODEs). This makes Stan a particularly attractive candidate tool for fitting deterministic and stochastic infectious disease models based on ordinary differential equations.

The main purpose of this paper is to explore how Stan could be used to fit mathematical models to infectious disease count data. In the Methods section, we provide a brief description of the most important features of Stan's implementation of HMC and VB so the reader can get familiar with the tools that Stan is based on. We then investigate three different examples and report our findings in the Results section. First, we consider a hierarchical model to infer age-specific gonorrhoea diagnosis rates while adjusting for spatial heterogeneity of Public Health regions in England. Next we consider dynamic models based upon systems of ODEs that describe transmission dynamics of a single or multiple influenza strains. Using single strain models we examine an outbreak of influenza at a British boarding school and we fit a multistrain model to UK influenza data from the 2017/18 season where even though the main strain curculating was a B strain, there was evidence of the H3 strain as well. The examples are presented using the R interface to Stan (rstan) and the rethinking R package~\cite[]{mcelreath2012rethinking}. The code is made freely available at  \url{https://github.com/anastasiachtz/COMMAND_stan.git}. 

\section{Material and methods}

\section*{Statistical inference using Stan}
\label{S:2}

Stan is an open-source general purpose inference software for a large range of Bayesian models, including regression, hierarchical models and state-space models. The software implements several numerical techniques for sampling from posterior distributions, most notably gradient-based sampling techniques, but also a method to approximate posterior distributions with variational inference, and penalized maximum likelihood estimation via numerical optimization. Gradient-based sampling is implemented through the No-U-turn Sampler(NUTS) \cite[]{hoffman2014no}, in combination with automatic differentiation to numerically approximate the gradients \cite[]{griewank1989automatic,griewank2008evaluating}. Variational inference aims to find an approximating probability distribution which is “close” to the posterior distribution of interest, and easy to sample from. It is implemented through stochastic optimization of a non-symmetric measure of the difference between the two distributions. Moreover, Stan provides a built-in mechanism for specifying and solving systems of ODEs, making it suitable for inference of SIR-type models.

Stan's probabilistic programming language is written in C++, with interfaces for R, Python, MATLAB, Julia, Stata, Mathematica, Scala and the command line. Users write Bayesian models in a computing language similar to standard statistical notation, much like the popular BUGS language. Detailed documentation is available, including User's Guide, Language Reference Manual and Functions Manual~\cite[]{Stan2018Stan}, as well as a separate guide for each of the Stan interfaces, all addressed to users of all experience levels. The User's Guide introduces readers incrementally to advanced modelling and programming techniques through a broad range of statistical models, and acts as a road map not only for learning Stan, but also modern Bayesian modelling in general. The Stan Language Reference Manual provides detailed analyses of the inference algorithms and clarifications on the Stan syntax. The Stan Functions Manual documents all integrated functions. 

Briefly, a difference between Stan and other automated platforms such as BUGS and JAGS, is that variable types and indices must be declared similarly as in the C++ programming language. Variables are declared by their type, in blocks according to their use, and constraints upon them need to be defined carefully. As seen in the example code in \hyperlink{appCοde}{Appendix B}, the first blocks of Stan's model statement consist of data, transformed data, parameters, transformed parameters and generated quantities. Within the model block, sampling notation is very similar to BUGS. User-defined probability functions can also be employed. The Stan code is written to a human-readable Stan model file, should have the extension .stan, and is portable across interfaces (e.g. R, Python, etc.) and operating systems (e.g. UNIX, Windows, Mac OS). According to the interface used, users need to call different functions for the different inference methods offered. All these functions include an argument which defines the location and name of the Stan model file. 

In the presence of missing data, inference is challenging in epidemic models. In Stan, missing continuous data can be treated as additional parameters, and thus are straightforward to handle. Users need to extend the Stan model file to identify which values are missing, and declare model parameters for each missing datum. However, with Stan, missing discrete data cannot be handled in the same manner due to the nature of the underlying inference algorithms. There is one notable workaround. When missing discrete data have a lower and upper bound, then it is possible to loop over all possible instances of missing values, sum the density value of the corresponding posterior distribution, and thus marginalize out the missing discrete data. The same process may, in principle, be applied to discrete bounded latent parameters.

The two main inference algorithms implemented in Stan are NUTS, the Hamiltonian Monte Carlo No U-Turn sampler, and Automatic Differentiation Variational Inference (ADVI) \cite[]{2015arXiv150603431K}. By changing just a few lines of code, it is possible to employ either of the algorithms, and also to build more complex mathematical models. In the following section we highlight the basic idea behind HMC-NUTS and ADVI as they are implemented in Stan.
A more detailed mathematical description of the algorithms is included in \hyperlink{appA}{Appendix A}.

\subsection{Hamiltonian Monte Carlo}

Statistical inference of epidemic models commonly rests on MCMC algorithms. These algorithms provide samples from the posterior probability distribution of model parameters by generating a Markov chain that has the target distribution, i.e. the posterior distribution of the model parameters, as its stationary distribution. The idea behind most MCMC techniques such as the Metropolis-Hasting algorithm \cite[]{metropolis1953equation,hastings1970monte} and Gibbs sampling \cite[]{geman1984stochastic} is to explore the parameter space by proposing a new sample based on the current sample and then accepting or rejecting it according to a certain probability. A frequent challenge is that the algorithm does not propose samples in regions of the parameter space that are distant from the current state. This may result in slow convergence to the stationary distribution when the parameter space with high posterior support is far from the initial values. It may also result in slow exploration of the parameter space with high posterior support when the target distribution has multiple distinct modes or an irregular shape \cite[]{hoffman2014no,neal1993probabilistic}. Thus, MCMC algorithms which take samples from a target distribution by making a random proposal and then accepting or rejecting it, may require very long run times, even though they are theoretically guaranteed to explore all the regions of the parameter space eventually.

In contrast to the Metropolis-Hastings and Gibbs sampling algorithms, HMC algorithms propose new samples adaptively, based on the gradients of the target distribution at the current state \cite[]{2012arXiv1206.1901N}. The theoretical foundation of HMC is based on concepts in differential geometry. Here we  sketch only the basic steps of HMC, see \cite{betancourt2017geometric} for a detailed exposition. First, the state space is augmented, adding to the parameters of interest auxiliary "momentum" parameters. Second, the Hamiltonian function, which is simply the negative log distribution of all the parameters, is formulated. The Hamiltonian function is associated with a physical interpretation, the total energy of a dynamic physical system in terms of object location and its momentum in time. The object's location relates to the potential energy and the momentum relates to the kinetic energy. Their sum, which is the total energy, defines the Hamiltonian. Third, the momentum parameters are sampled, typically from some Gaussian distributions, given the current values of the parameters of interest. Fourth, the proposal distribution of the parameters of interest is constructed conditional on the gradients of the Hamiltonian at the current value and thus takes into account the local geometry of the distribution. 

Most HMC implementations, including that in Stan, are based on the leapfrog method to construct the proposal density. The method alternates between half-step updates of the momentum parameters and full steps of the parameters of interest~\cite[]{beskos2013optimal}. The gradients of the posterior distribution are typically not known analytically, and so they are numerically approximated. Stan uses automatic differentiation\footnote{Automatic differentiation, instead of computing the expressions of the derivatives, decomposes the complex expressions into primitive ones and computes the derivatives through accumulation of values during code execution, resulting in numerical derivatives.} for this sub-task \cite[]{griewank2008evaluating,2015arXiv150907164C}. An accept-reject step ensures that the resulting samples are asymptotically from the target distribution.

The standard HMC algorithm has a number of tuning variables, that complicate automated numerical inference \cite[]{2016arXiv160100225B,2014arXiv1411.6669B}. These include the number of leapfrog steps i.e. the number of updates performed before acceptation or rejection, the length of each update (following the gradient), and the covariance matrix of the probability distribution of the momentum parameters. In Stan, an adaptive version of the leapfrog algorithm is implemented in order to reduce the number of tuning variables. The covariance matrix of the momentum parameters is estimated during warm-up, as is the step size, aiming at a specific target acceptance rate~\cite[]{Stan2018Stan}. 
The optimal number of updates is determined dynamically. The idea is to use a sufficient number of update steps to explore the parameter space in an efficient manner. This is achieved by either avoiding a U-turn to previously explored trajectories or stopping at a predetermined maximal number of increasing the leapfrog steps. Stan's NUTS algorithm uses multinomial sampling from each trajectory to select a sample~\cite[]{Stan2018Stan,2017arXiv170102434B,hoffman2014no}. If the leapfrog integrator fails in the sense that the value of the Hamiltonian is far from its initial value, then the designed trajectory is identified as divergent and rejected.

HMC requires more computational effort at every step compared to standard MCMC techniques, primarily because of the gradient calculations. However, this feature enables HMC algorithms to explore target distributions of highly correlated parameters more effectively than standard MCMC. This implies that much fewer iterations are typically needed to estimate model parameters and their uncertainty intervals, and therefore that the overall computational runtime of HMC algorithms can be substantially less compared to standard MCMC techniques. In particular,  \cite{monnahan2017faster} demonstrate that over a range of examples, Stan-based HMC typically returns a higher effective sample size per computational unit compared to MCMC as implemented in JAGS.

\subsection{Variational Inference}
There are real-life applications in statistics where we cannot easily use the MCMC approach due to time constraints, as is the case e.g. for real-time inferences when managing outbreaks of emerging pathogens. In these cases, we may be willing to partially sacrifice accuracy for computational speed. Variational inference is a method which originates from machine learning and tends to be faster than MCMC~\cite[]{jordan1999introduction,wainwright2008graphical}. 

At its core, variational inference relies on translating the problem of directly estimating posterior distributions into an optimization problem that aims to find an easy-to-compute density that is close to the posterior. More formally, variational inference considers a family of approximating distributions to the posterior distribution. Each member of this family is a candidate approximating density to the posterior density. The goal is to find the closest candidate in terms of the Kullback-Leibler (KL) divergence to the exact density~\cite[]{blei2017variational}. The KL divergence is essentially a measure of the information lost when the candidate density is used to approximate the exact posterior~\cite[]{kullback1997information}. It is expressed as the expectation, with respect to the approximation, of the difference between the log approximating distribution and the log posterior distribution given the data. In other words the KL divergence is a non-symmetric measure of the difference between the two probability distributions. Since the KL divergence involves the posterior, it is not computable. Consequently, variational inference maximizes a proxy to the KL divergence, the Evidence Lower Bound (ELBO), which is equivalent to the KL divergence up to a constant (see \hyperlink{appA}{Appendix A})

In Stan, the automatic differentiation variational inference (ADVI) method is implemented. The fact that we need to optimize the KL divergence implies a constraint that the support of the chosen approximation lies within the support of the posterior \cite[]{2015arXiv150603431K}. However finding such a family of approximating densities is very difficult. To overcome this challenge, ADVI transforms the support of the parameters of interest to the real coordinate space, ensuring that the aforementioned constraint is always valid. Then, all parameters are defined on the same space so that we can choose the variational approximation independent of the model. To this end, Stan provides a library of transformations. Considering then a Gaussian variational approximation on the transformed space, ADVI tries to maximize the ELBO. Note that the variational approximation in the original parameter space is non-Gaussian and its shape is directly determined by the form of the transformation used.

Stan offers two options for the Gaussian approximation used. The first is mean-field ADVI, which simply assumes that the unknown parameters are independent. Mean-field variational Bayes is widely used since it is fast, however there is no theoretical guarantee for accurate results~\cite[]{wang2018frequentist}. An additional concern is that the marginal variances of the parameters are often under-estimated~\cite[]{turner2011two, bishop2006pattern}. The second option is full-rank ADVI.  This approach dispenses with the independence assumption that underlies mean-field variational Bayes, and is therefore theoretically superior in capturing posterior correlations ~\cite[]{wang2018frequentist}. However full-rank ADVI can be challenging to implement in practice.

In contrast to standard variational inference algorithms that maximize ELBO using coordinate ascent, ADVI uses a gradient-based algorithm to perform the maximization. In particular, ADVI is based on a stochastic gradient ascent algorithm where the gradients are computed using automatic differentiation ~\cite[]{2015arXiv150603431K}. Despite the fact that ADVI in Stan is a faster alternative to MCMC and is automated in the sense that the user needs to provide only the model and the data, it may fail for several reasons. As in every variational inference approach, initialization plays a crucial role and we can only test random initializations. Also, the fact that the posterior in the transformed space may not be well-approximated by a multivariate normal  or that this specific iterative algorithm may not be able to find that optimal multivariate normal, may lead to poor performance.

\section*{Modelling}

\label{S:3}

\subsection{Bayesian multi-level models}
Heterogeneity is pervasive in epidemiology, including for example heterogeneous patient groups, heterogeneous treatment effects in different locations, or heterogeneous time effects. Statistically, Bayesian multi-level models are the basic modeling tool in these cases, and well suited to making inferences from structured data sets~\cite[]{gelman2006data}. Stan was originally designed as a general-purpose platform for Bayesian inference for multilevel models while trying to overcome difficulties arising from using BUGS or JAGS~\cite[]{lunn2012bugs,plummer2003jags,Stan2018Stan}, and so this will be our first example. We provide an example of estimating gonorrhea diagnosis rates in the context of heterogeneity across age groups, gender and Public Health regions in England. 

Data on gonorrhoea case counts were obtained from Public Health
England, \url{https://www.gov.uk/government/statistics/sexually-transmitted-infections-stis-annual-data-tables}. The data we use here range from 2012 to 2016 and are stratified by gender ($m=0$ for female, $m=1$ for male), age group ($a= 0,\dotsc,6$ for the age categories years $13-14$, $15-19$, $20-24$, $25-34$, $35-44$, $45-64$ and $65+$), and PHE region ($r=1, \dotsc, 9$ for East Midlands, East of England, London, North East, North West, South East, South West, West Midlands, Yorkshire \& the Humber). Population denominators for each group are available from the same source, and denoted by $P_{ram}$. 
\vspace{-.27cm}
\begin{figure}[H]
  \centering
  \includegraphics[width=\linewidth]{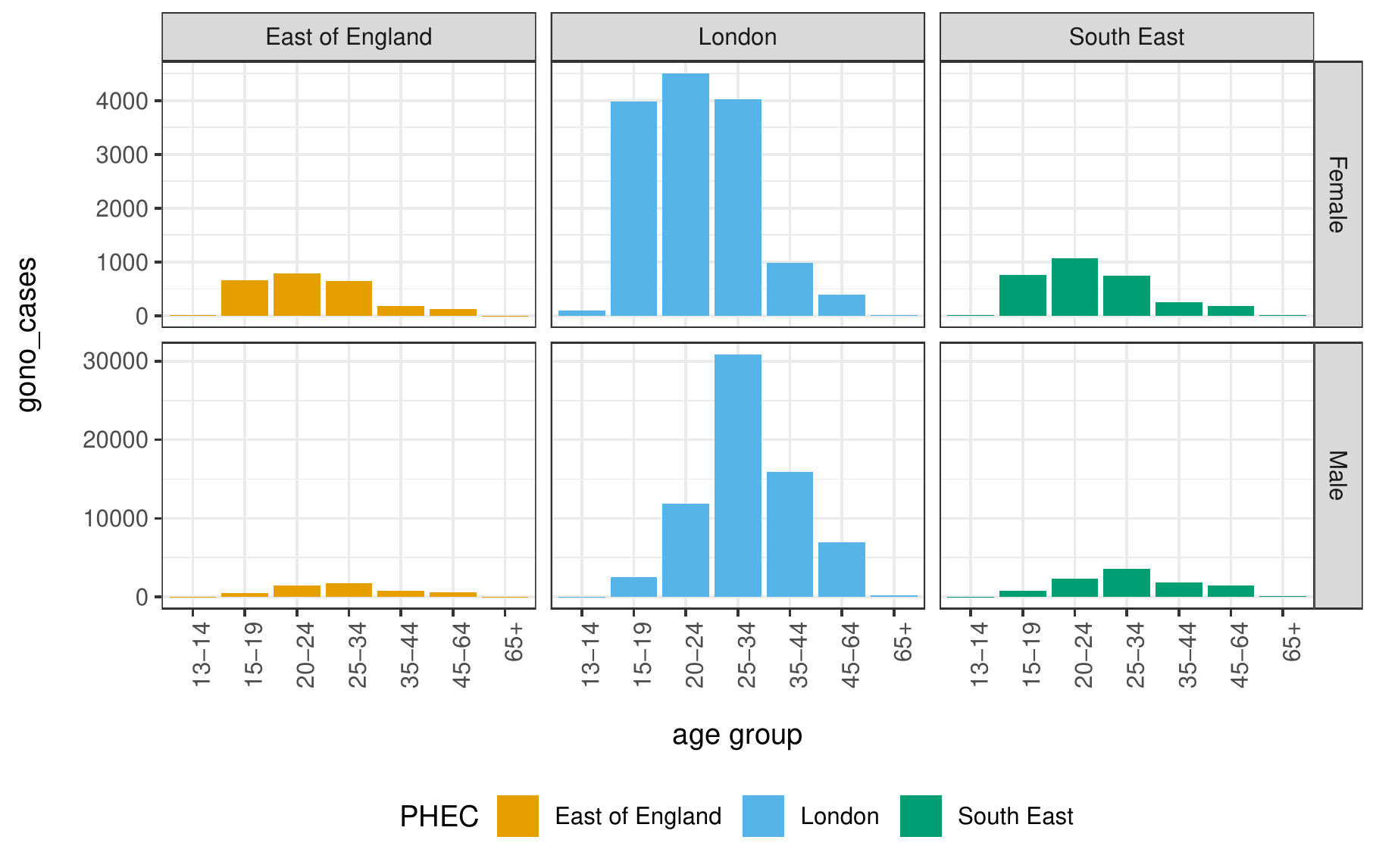}
  \caption{\label{fig:gono_data}\scriptsize{Gonorrhoea case counts in England. The total number of reported cases between 2012 and 2016 are shown by age (x-axis), gender (rows) and three of the Public Health England regions (columns), namely East of England, London and South East. For visualisation purposes, different limits on the y-axis were chosen for men and women. There were substantially more reported cases among men.}}
\end{figure}

Figure \ref{fig:gono_data} illustrates the substantial variation in the number of diagnoses by age, gender, and location. Specifically, we note that diagnoses peak at younger ages among women when compared to men, which can be modelled through separate age-specific random effects. Further, we notice that diagnoses among males from London are substantially higher and since the sample size in London is large, this is unlikely due to error. So, if this is not accounted for, the overall estimates will be biased upwards, suggesting to add an independent effect for London men to the model. A typical approach for estimating region-, age- and gender-adjusted standardised diagnosis rates is via Poisson multi-level models, for example

\begin{equation}\label{eq:ref1}
\begin{split}
Y_{ram}  \sim &\:\operatorname{Poisson} \left(\kappa_{ram}\right)\\
log(\kappa_{ram})=&\:\alpha+\alpha_{r}+log(\mathrm {P}_{ram})+\\ 
&\quad\xi_{a}\hspace{.1cm} \mathrm {M}_{ram} + \nu_{a}\hspace{.1cm}(1- \mathrm{M}_{ram})  +\\ 
&\quad\beta_{M}\hspace{.1cm}\mathrm {M}_{ram}+ \beta_{ML}\hspace{.1cm}\mathrm {M}_{ram}\hspace{.1cm} \mathrm{L}_{ram}\\
\alpha \sim &\:\mathcal{N} (0,100)\\
\beta_{M}\sim & \:\mathcal{N} (0,10)\\ \beta_{ML}\sim & \:\mathcal{N} (0,10)\\
\alpha_{r}\sim&\:\mathcal{N} (0,\sigma_{\alpha}^2)\\
\xi_{a}\sim & \:\mathcal{N} (0,\sigma_{\xi}^2)\\
\nu_{a}\sim & \:\mathcal{N} (0,\sigma_{\nu}^2)\\
\sigma_{\alpha}^2 \sim &\: Exp(1)\\
\sigma_{\xi}^2 \sim &\: Exp(1)\\
\sigma_{\nu}^2 \sim &\: Exp(1).
\end{split}
\end{equation}
In the above, $\mathrm{Y}_{ram}$ are the number of gonorrhoea cases per strata, $\mathrm {M}_{ram}$ a gender indicator variable (0 for female, 1 for male) and $\mathrm {L}_{ram}$ a location indicator variable (0 for outside London, 1 for London). The model includes a baseline term ($\alpha$), a fixed gender effect ($\beta_{M}$), a fixed interaction effect between gender and location ($\beta_{ML}$), region-specific random effects  ($\alpha_r$), age-specific random effects among men ($\xi_a$), and age-specific random effects among women ($\nu_a$). In total, there are 29 parameters to estimate. 

\subsection{Deterministic ODE-based models}
The dynamics of disease spread are frequently formulated in terms of ODE-based models, in whom the study population is divided into compartments representing a specific stage of the epidemic or a demographic status, such as susceptible, infected, and recovered individuals~\cite[]{kermack1927contribution, anderson1992infectious}. The disease dynamics are captured in a system of non-linear ODEs, such as the susceptible-infectious-recovered (SIR) model:
\begin{align}\begin{split}
\frac{\mathrm{d}S}{\mathrm{d}t} & = - \beta \frac{I(t)}{N} S(t)\\
\frac{\mathrm{d}I}{\mathrm{d}t} & = \beta \frac{I(t)}{N} S(t) - \gamma I(t)\\
\frac{\mathrm{d}R}{\mathrm{d}t} & = \gamma I(t)
\end{split}\end{align}
where $S(t)$ represents the number of susceptible, $I(t)$ the number of infected and $R(t)$ the number of recovered individuals at time t. The total population size is denoted by $N$ (with $N = S(t) + I(t) + R(t)$), $\beta$ denotes the transmission rate and $\gamma$ denotes the recovery rate. In an outbreak scenario, typical initial conditions are $I(0)=1$, $S(0)=N-1$ and $R(0)=0$. 

We usually want to obtain estimates of $\beta$ and $\gamma$, the basic reproduction number $R_{0}$ which is defined as $\beta/\gamma$ for the SIR model, and the initial number of susceptible individuals. The data typically consists of the number of new infections within a certain time interval, such as days or weeks. Inference is then complicated by the fact that the model states $S$, $I$, $R$ are typically latent variables, and by the non-linear nature of the disease dynamics.

Stan has two built-in ODE solvers, which enable inference of a variety of ODE-based models. The first solver is for non-stiff dynamic systems, i.e. systems whose components evolve at similar rates, is based on the fourth and fifth order Runge-Kutta method, and fast. The second solver is for stiff systems, i.e. systems consisting of components that evolve at different time scales, is slower, and more robust~\cite[]{Stan2018Stan}.

In what follows, we provide a setting where the ODE solver role is highlighted in the context of a deterministic SIR model. We examine an outbreak of influenza A (H1N1) at a British boarding school in 1978 . The data consist of daily counts $Y_{t}$ of the number of infected students, over a time interval of 14 days. To link the data to the SIR dynamics, we can specify the following Poisson observation model:
\begin{equation}
\label{eq:ref5}
    Y_{t}\sim \operatorname{Poisson} \left({\lambda_{t}}\right)
\end{equation}
\begin{equation}
\label{eq:ref6}
    \lambda_t=\int_{0}^{t}\left({\beta \frac{I(s)}{N} S(s) - \gamma I(s)}\right) \mathrm{d}s
\end{equation}

We aim to estimate $\beta$, $\gamma$, the initial proportion of susceptible individuals $s(0)$, and implicitly the initial proportion of infected individuals $i(0)$ (assuming that the initial proportion of removed individuals is 0 then $i(0)=1-s(0)$).
To do so, we specify the following priors
\begin{equation}\label{eq:pr_SIR}
\begin{split}
\beta\sim \operatorname{Log normal} (0,1)\\
\gamma\sim \Gamma (0.004,0.002)\\
s(0)\sim Beta (0.5,0.5).
\end{split}
\end{equation}

\subsection{Stochastic ODE-based models}
Even though the deterministic approach gives us an insight into the dynamics of the disease, considering demographic stochasticity may allow for a more accurate estimation of the parameters related to the spread of the disease, as the stochastic component can absorb the noise generated by a possible mis-specification of the model~\cite[]{andersson2012stochastic,malesios2017bayesian}. A natural way to do so in the above Poisson model is via employing the continuous-time analog of the auto-regressive (1) model, the Ornstein-Uhlenbeck (OU) process~\cite[]{karatzas1998brownian} as follows:
\begin{equation}
    \textnormal{ $Y_{t}\sim \operatorname{Poisson} \left({\lambda_{t}}\right)$}\\
\end{equation}
\begin{equation}
    \textnormal{ $\lambda_{t}=exp(\kappa_{t})$}
\end{equation}
\begin{equation}
    \textnormal{ $\mathrm{d}\kappa_s = \phi (\mu_t - \kappa_s)\mathrm{d}t + \sigma\mathrm{d}B_s $}
\end{equation}
where $B_s$ denotes standard Brownian motion, $\sigma$ is the instantaneous
diffusion term, $\phi$ is the speed of reversion of $\kappa_t$ and $\mu_t$ is a piecewise constant function which corresponds to the logarithm of the solution of the deterministic model:
\begin{equation}
    \mu_t=\log\left({\int_{0}^{t}\left({\beta \frac{I(s)}{N} S(s) - \gamma I(s)}\right) \mathrm{d}s}\right)
\end{equation}
The instantaneous $\kappa_{t}$ is an OU process evolving around $\mu_t$. Its transition density from day $t$ to day $t+1$ is available in closed form:
\begin{equation}
    \kappa_{t+1}|\kappa_t\sim \mathcal{N} \left({\mu_t + (\kappa_t - \mu_t)e^{-\phi},\frac{\sigma^2}{2\phi}(1-e^{-2\phi})}\right).
\end{equation}
To complete the model specification, we considered a half-normal prior distribution for $\phi$ with large variance, $\phi\sim \operatorname{Half Normal} (0,100)$ and an inverse-gamma prior density for $\sigma^2$, $\sigma^2\sim \operatorname{Inv-Gamma} (0.1,0.1)$.

\subsection{Multistrain models}
Lastly we explore fitting ODE-based multistrain models with Stan. Specifically we focus on a multistrain SIR model in which  each strain acts independently:  
\begin{align}
\begin{split}
\frac{\mathrm{d}S_x}{\mathrm{d}t} & = - \beta_x \frac{I_x(t)}{N} S_x(t)\\
\frac{\mathrm{d}I_x}{\mathrm{d}t} & = \beta_x \frac{I_x(t)}{N} S_x(t) - \gamma I_x(t)\\
\frac{\mathrm{d}R_x}{\mathrm{d}t} & = \gamma I_x(t),\\
\end{split}
\end{align}
where \(S_x(t)\) denotes the number of susceptibles to strain \(x\) at time \(t\)
and similarly \(I_x(t)\) and \(R_x(t)\)
denote the number of infected and recovered individuals to strain \(x\) at time \(t\). The model consists of overlapping compartments, with total population size
(\(N=S_x(t) + I_x(t) + R_x(t)\)) for any strain \(x\). \(\beta_x\) is the strain-specific transmission rate and \(\gamma\) is the recovery rate, modelled as identical for each strain.

The model is fitted to weekly influenza-like illness (ILI) case counts, and virological data. To fit the model to the data, we track the number of ILI cases due to strain \(x\), denoted by \(\mathrm{ILI}^{+,x}(t)\), as well as the number of ILI cases that are not a result of infection with any of the influenza strains, denoted by \(\mathrm{ILI}^{-}(t)\). The total number of ILI cases is then:
\(\mathrm{ILI}(t) = \sum_x \mathrm{ILI}^{+,x}(t) + \mathrm{ ILI}^{-}(t)\).

The cumulative number of ILI cases over time is modelled as follows:
\begin{align}
\begin{split}
\frac{\mathrm{dILI}^{+,x}}{\mathrm{d}t} & =  \theta_x^+\beta_x \frac{I_x(t)}{N} S_x(t)  - \mathrm{ILI}^{+,x}(t) \delta(t \bmod 7) \\
\frac{\mathrm{dILI}^{-}}{\mathrm{d}t} & =  \theta^-(t)(N - \sum_x I_x(t)) - \mathrm{ILI}^{-}(t) \delta(t \bmod 7),
\end{split}
\end{align} 
where \(\theta_x^+\) denotes the probability of symptomatic ILI infection, and \(\theta^-\) the probability of developing ILI symptoms when not having flu. \(\delta(t \bmod 7)\) is the Dirac delta function, which integrates to $1$ when \(t \bmod 7 = 0\), i.e. at the start of every week, and is otherwise $0$. These equations model cumulative ILI incidence over time, while being reduced to zero at the beginning of the week (due to the Dirac delta function). This is in line with the data, which counts the cumulative number per week (i.e. restarts counting at zero every week).

It is well known that flu-negative ILI rates increase in winter \cite[]{fleming2008lessons}. To account for this, we modelled \(\theta^-\) to change over time via
\[
\log\theta^-(t) = \hat{\theta} + \phi\left(e^{-\frac{(t-\mu_t)^2}{2 \sigma^2}}-1\right),
\]
where \(\hat{\theta}\) is the maximum value of the (log) value of the
flu negative ILI rate, \(\phi\) is the amplitude of the peak, \(\mu_t\)
is the time of the peak and \(\sigma\) governs the width of the peak.

We have now everything in place to link the multi-strain model to the data through the variables \(ILI^{+,x}\) and \(ILI^{-}\). First we assume that the number of ILI cases visiting a GP follows a binomial distribution $\mathcal{B}$ such that the likelihood $\mathcal{L}$ of the model outcomes and parameters given the number of ILI diagnoses per week can be defined as follows:
\[
\mathcal{L}(\mathrm{ ILI}, \epsilon; y^{\mathrm{ ILI}}, N, N_c) = \mathcal{B}(y^{\mathrm{ ILI}}; \mathrm{ ILI} N_c/N, \epsilon),
\]
where \(y^{\mathrm{ILI}}_i\) is the observed number of ILI cases in the
monitored population \(N_c\), \(N\) is the total population,
\(\mathrm{ILI}\) is the total predicted ILI cases in the population (see
above) and \(\epsilon\) is the rate with which someone with ILI is
diagnosed, i.e. this is a combination of the probability that a
symptomatic (ILI) case consults the GP and the GP correctly diagnosing the patient. Note that the number of ILI cases in the population is scaled to the expected number of ILI cases in the monitored population using $\mathrm{ILI} N_c/N$.

The virological samples are assumed to follow a multinomial
distribution:
\[\mathcal{M}(y^{+,x_0},\dots, y^-; \mathrm{ ILI^{+,x_0}}/\mathrm{ ILI}, \dots \mathrm{ ILI^{-}}/\mathrm{ ILI})\]
where \(y^{+,H1}, y^{+,H3}, y^{+,B}\) represent the number of positive samples for
each strain, \(y^-\) is the number of negative samples and
\(\mathrm{ ILI^{+,x_0}}/\mathrm{ ILI}, \dots, \mathrm{ ILI^{-}}/\mathrm{ ILI}\)
are respectively the probability of finding positive samples with each
strain \(x_0, \dots\) and finding negative samples (flu negative ILI).\\

\section{Results}
\label{S:4}
\subsection{Poisson multi-level model}

We fit our full hierarchical model (\ref{eq:ref1}) using Stan's NUTS algorithm. First of all, we tested for convergence to the target distribution, by inspecting the trace plots of multiple chains that were started from distinct initial values. Next, we tested for sufficient exploration of the target distribution, by calculating effective sample sizes for each model parameter, which are an estimate of the number of independent draws from the marginal posterior distributions that are represented in the numerical output. Using R, effective sample sizes can be computed through the bayesplot or coda packages, see \hyperlink{appCοde}{Appendix B}. Here, to obtain effective sample sizes above 500, approximately 30,000 iterations are needed. This is pretty good, with no further tuning required. Computations took us about 13 minutes.

\begin{figure}
     \centering
     \begin{subfigure}[b]{.85\linewidth}
         \centering
         \includegraphics[width=\linewidth,keepaspectratio]{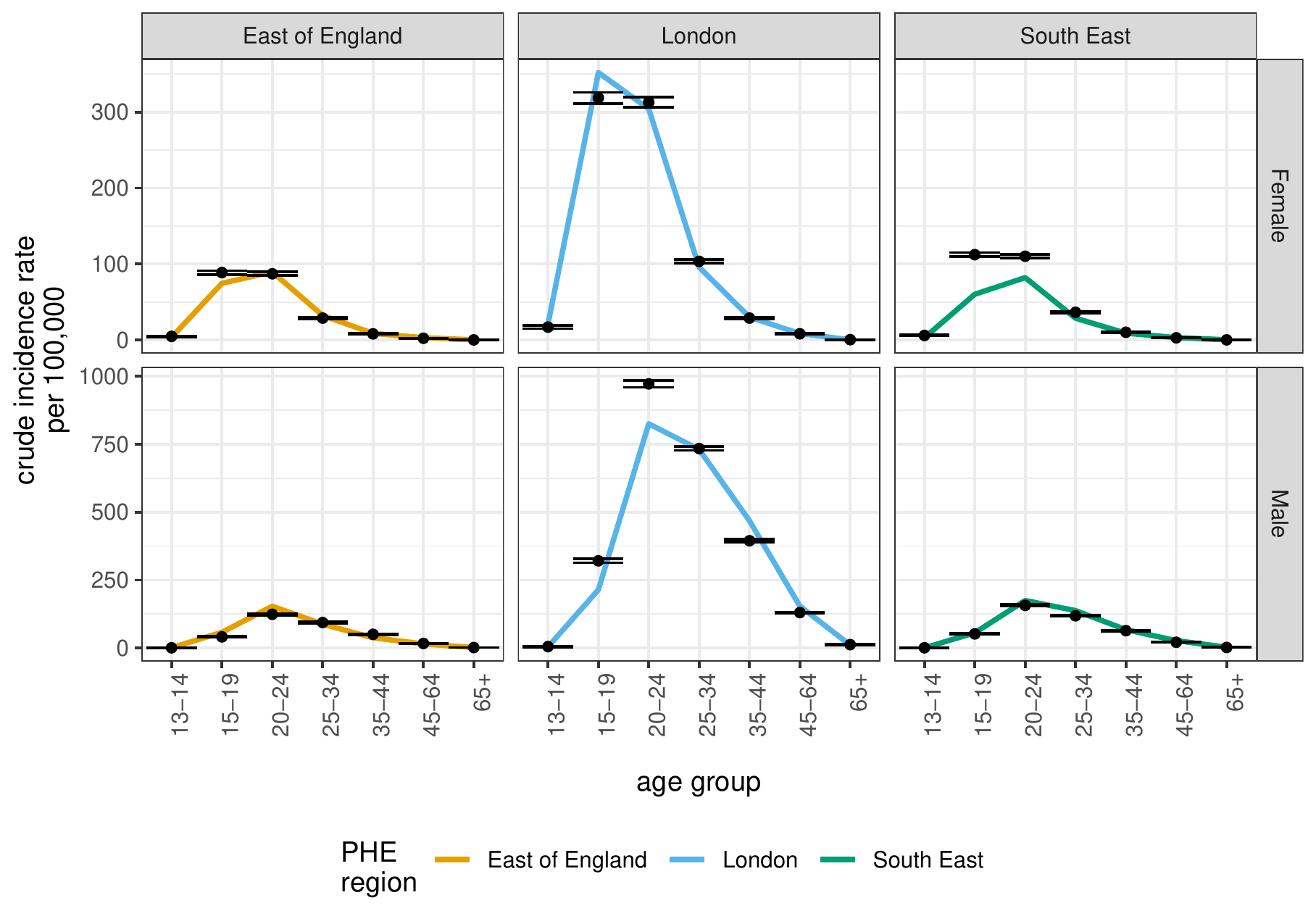}
         \caption{\label{fig:gon_fit}Region-, age- and gender-specific posterior estimates of gonorrhoea diagnosis rate. Posterior medians(black dots), 95$\%$ credibility intervals(black error bars) and data(colored lines).}
     \end{subfigure}
     \hfill
     \begin{subfigure}[b]{.85\linewidth}
         \centering
         \includegraphics[width=\linewidth,keepaspectratio]{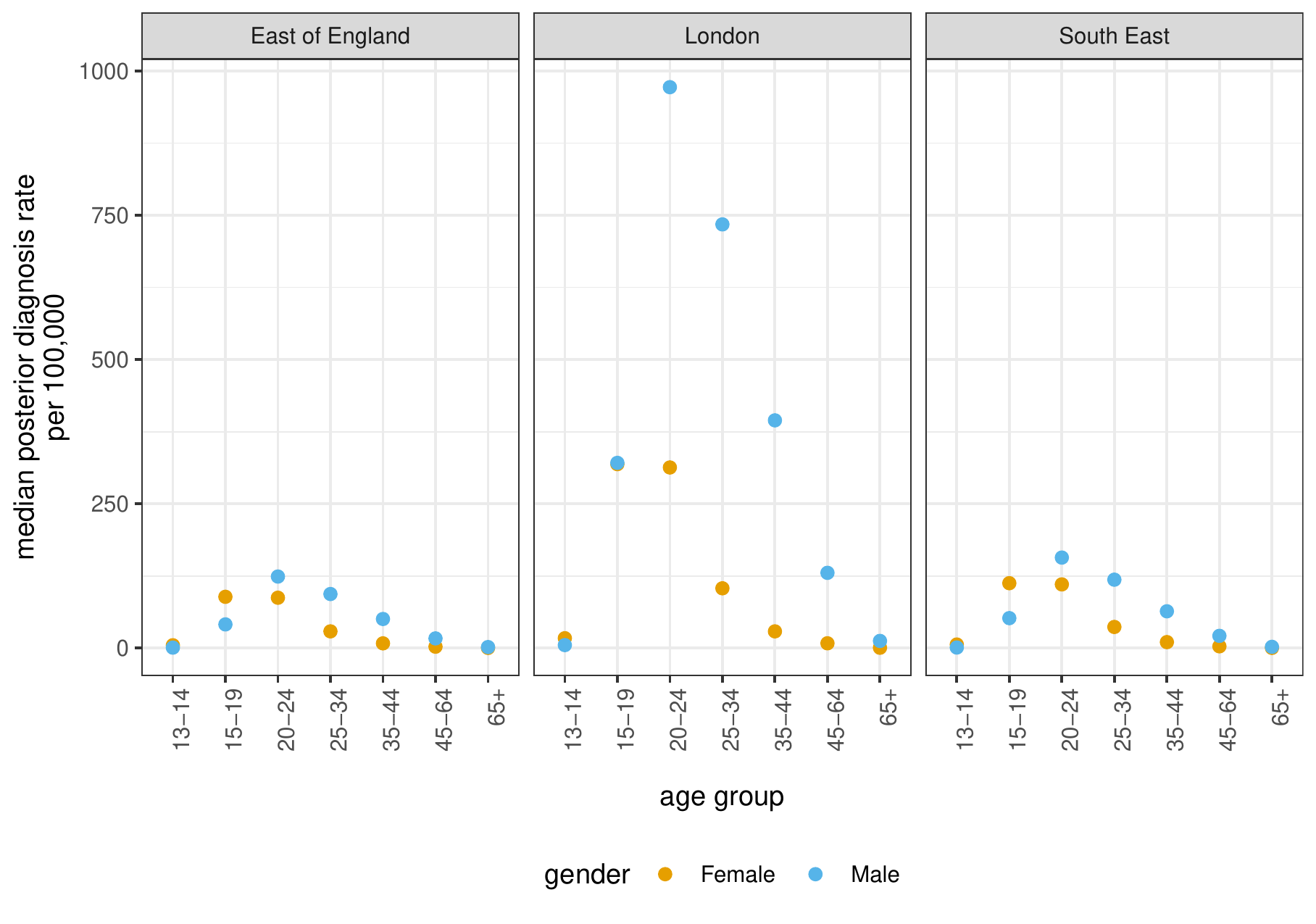}
         \caption{\label{fig:gon_fit_gender}Comparison of gonorrhoea diagnosis rate estimates between males and females.}
     \end{subfigure}
        \caption{Inference results for gonorrhoea hierarchical model using NUTS}
\end{figure}

Figure \ref{fig:gon_fit} illustrates the region-, age- and gender-specific posterior estimates of standardised gonorrhea diagnoses rates per 100,000 individuals (black dots and error bars). Adding crude diagnosis rate estimates (colored lines), it can be seen that the model achieves an overall reasonable fit, which could be further assessed through posterior predictive checks. As suggested in figure \ref{fig:gon_fit_gender}, the model indicates further that young women aged 15-19 have higher risk of acquiring gonorrhoea than their male peers. In contrast, among age groups 20-64, men have higher risk of acquiring gonorrhoea when compared to their female peers. However the model fit also reveals notable regional trends. For example, in the South East and South West, the model substantially overestimates disease risk among young women aged 15-24. This suggests that in these regions, diagnoses rates among young women aged 15-24 are lower than expected under the general trends captured in model (\ref{eq:ref1}). Alternative explanations could relate to biases in data collection.

\subsection{Single strain SIR models}

In 1978, there was a report to the British Medical Journal for an influenza outbreak in a boarding school in the north of England. There were 763 male students which were mostly full boarders and 512 of them became ill. The outbreak lasted from the 22nd of January to the 4th of February and it is reported that one infected boy started the epidemic and then it spread rapidly in the relatively closed community of the boarding school. We use the data from Chapter 9 of \citet{de2006course} which are freely available in the R package outbreaks, maintained as part of the R Epidemics Consortium (RECON; \url{http://www.repidemicsconsortium.org}). Data consist of the number of students who are confined to bed each day which we assume that is equal to the total number of infected students each day.
 
Both models are fitted using Stan's NUTS algorithm using 5 chains, each with 100500 iterations of which the first 500 are warm-up to automatically tune the sampler, and then a sample is saved every fiftieth samples, leading to a total of 10000 posterior samples. We examine the convergence of the parameters by inspecting the trace plots of all chains indicating that there is no lack of convergence for both models and by checking the $\hat{R}$ convergence statistic reported by Stan. Ιf the chains have not yet converged to a common distribution the $\hat{R}$ statistic will be greater than one \cite[]{gelman2013bayesian, Stan2018Stan}. However, if it is equal to 1, it does not necessarily indicate convergence. As all convergence diagnostics, $\hat{R}$ can only detect failure to convergence but it cannot guarantee convergence. In our example, all models show good mixing according to the effective sample size, $\hat{R}$ and the trace plots. 

We also fit the models using the mean-field ADVI variant of Stan. All models were sensitive to initial values so we initialize our parameters using values drawn uniformly from the credible intervals we obtain from NUTS. In our example, the full-rank variant was not feasible, maybe due to the fact that observing only 14 days throughout the outbreak does not give us enough information to estimate the possible correlations.

\begin{figure}
     \centering
     \begin{subfigure}[b]{.92\linewidth}
         \centering
         \includegraphics[width=\linewidth]{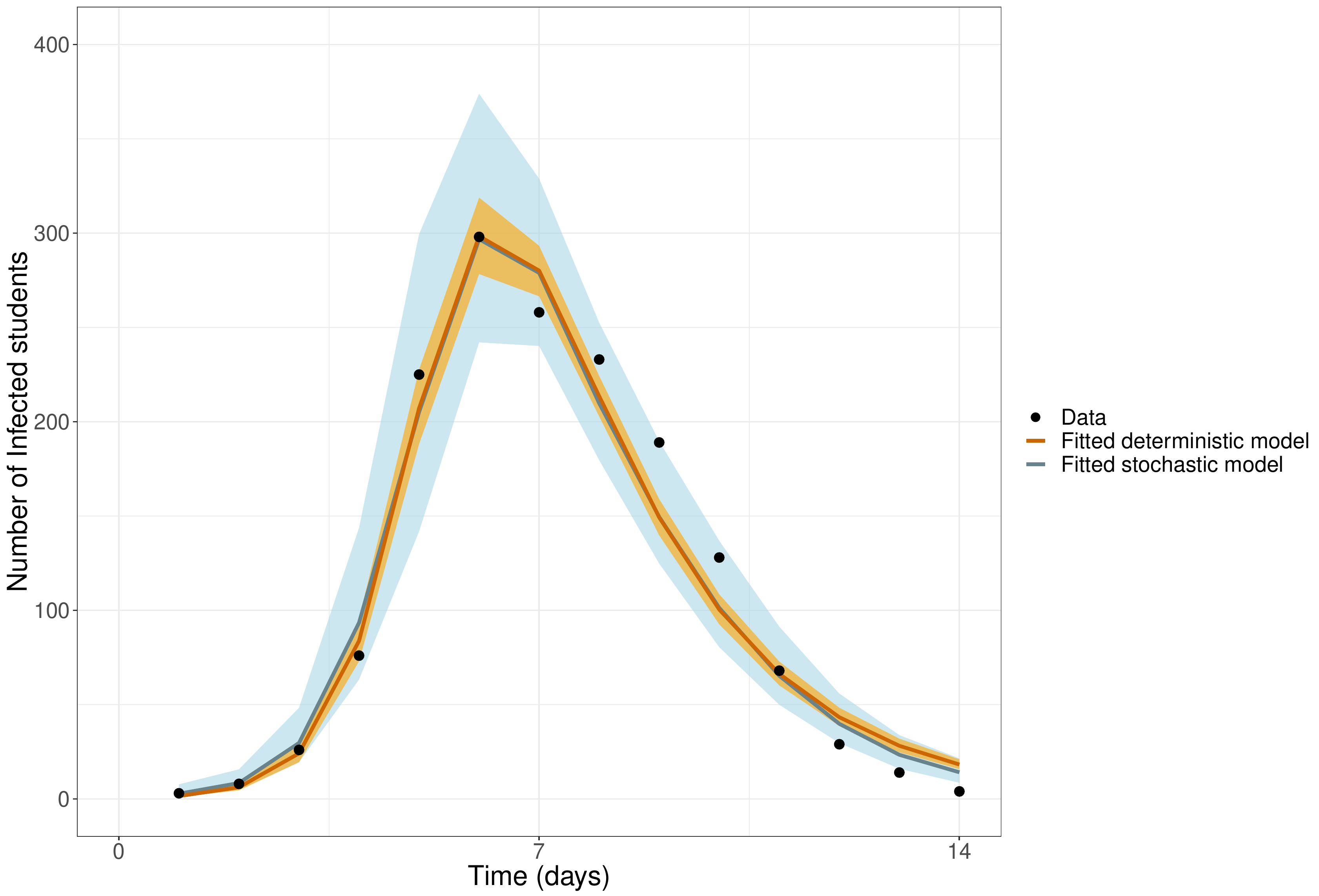}
         \caption{\label{fig:HMC_fit}NUTS}
     \end{subfigure}
     \hfill
     \begin{subfigure}[b]{.92\linewidth}
         \centering
         \includegraphics[width=\linewidth]{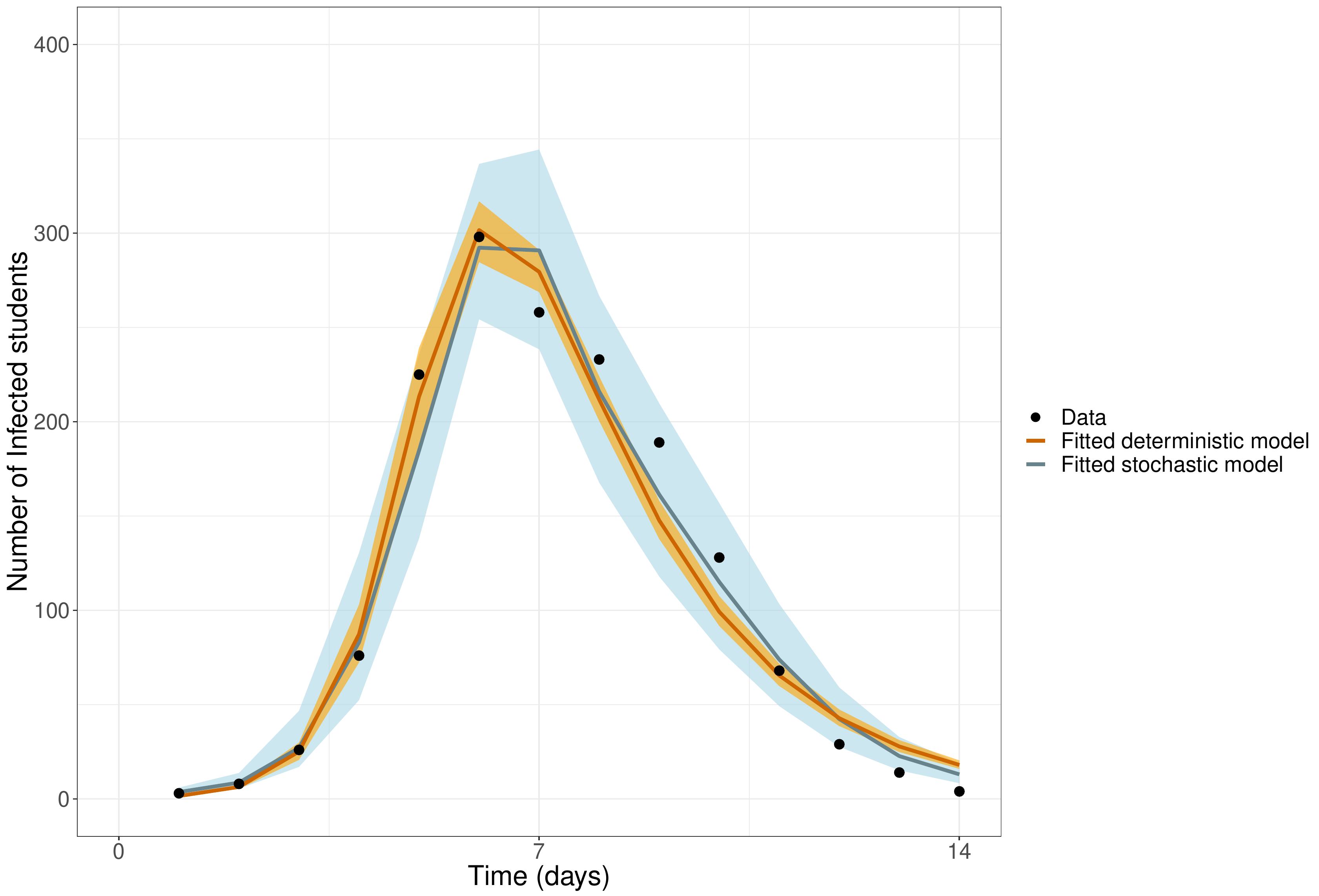}
         \caption{\label{fig:ADVI_fit}ADVI}
     \end{subfigure}
        \caption{\label{fig:single_fit}Inference results using NUTS and ADVI for influenza outbreak in British boarding school. Fit of the deterministic and stochastic SIR model to the data(black dots). Medians(lines) and 95\% CI(shaded areas)}
\end{figure}

For both the deterministic and the stochastic setting, posterior medians and 95$\%$ credible intervals of the parameters are summarized in Table \ref{table:model_1_2}. In both models, ADVI results in narrower credible intervals for $\beta$ and the basic reproduction number $R_0$ compared to NUTS, suggesting that ADVI may be underestimating the posterior uncertainty, as has been observed in the past. In general, the posterior estimates for $R_0$ are in line with the estimated $R_0$ obtained by \citet[]{wearing2005appropriate}. As seen from Figure \ref{fig:HMC_fit} and \ref{fig:ADVI_fit}, the deterministic model has a reasonable fit to the data but underestimates the overall uncertainty thus resulting in overly precise estimates which fail to capture the data appropriately.

Results from the stochastic model as summarized in Table \ref{table:model_1_2}, include additionally the parameters characterizing the transmission dynamics of the disease, so we also report posterior estimates for the parameter $\phi$ of the OU process which reflects the speed of reversion and the instantaneous variance $\sigma^2$. Again, the resulting 95$\%$ credible intervals from ADVI have shorter length compared to NUTS. 

\begin{table}[H]
 \caption{\label{table:model_1_2}HMC-NUTS using 5 chains, each with iter=100500; warmup=500; thin=50; 
post-warmup draws per chain=2000, total post-warmup draws=10000 ; ADVI(mean-field) using iter=10000, tol\_rel\_obj=0.01}
    \begin{adjustbox}{width=\textwidth}
    \centering
    \begin{tabular}{c|ccc|cc||ccc|cc}
      & \multicolumn{5}{c||}{Single Strain Deterministic model} & \multicolumn{5}{c}{Single Strain Stochastic model} \\ \hline
      & \multicolumn{3}{c|}{HMC} & \multicolumn{2}{c||}{ADVI} & \multicolumn{3}{c|}{HMC} & \multicolumn{2}{c}{ADVI} \\ \hline
            & mean & 95\% CI   & ESS  & mean  & 95\% CI     & mean    & 95\% CI    & ESS    & mean    & 95\% CI \\ \hline
        $\beta$ & 1.89 & 1.78-2.00 & 9766 & 1.89   & 1.86-1.93   & 2.02   & 1.68-2.71      & 9824     & 2.02      & 1.85-2.21 \\
    $\gamma$ & 0.48 & 0.46-0.50 & 10093 & 0.48   & 0.46-0.50   & 0.53   & 0.44-0.65      & 9965     & 0.55      & 0.45-0.66\\
    $s(0)$   & 1.00 & 1.00-1.00 & 9632 & 1.00  & 1.00-1.00   & 1.00   & 1.00-1.00      & 9034     & 1.00      & 1.00-1.00 \\
    $R_0$     & 3.93 & 3.67-4.22 & 9667 & 3.96   & 3.77-4.16  & 3.84   & 2.80-5.79      & 9976     & 3.73      & 2.98-4.60 \\
    $\phi$    &      &           &       &     &             & 4.34   & 0.46-19.19     & 9196     & 0.86      & 0.58-1.26  \\ 
   $\sigma^2$ &      &           &       &     &             & 2.63   & 0.36-12.32     & 8599     & 0.70      & 0.45-1.02 
    \end{tabular}
    \end{adjustbox}
\end{table}

\begin{table}[H]
\centering
\caption{\label{table:time}Execution time (minutes)}
   \begin{adjustbox}{width=0.7\textwidth}
   \centering
\begin{tabular}{c|c|c}
    & Single Strain Deterministic model  & Single Strain Stochastic model      \\ \hline
HMC & 13.63     & 47.68  \\
ADVI  & 0.32     & 1.86 
\end{tabular}
\end{adjustbox}
\end{table}

Summing up, the results of both the deterministic and the stochastic setting bring us to the preliminary conclusion that if we are interested in real-time inference both methods are feasible and efficient. In terms of computational time ADVI is extremely efficient (Table \ref{table:time}). As Figure \ref{fig:single_fit} demonstrates, adding stochasticity improves the fit to the data.

\subsection{Multistrain model}

\begin{figure}[H]
\centering
\includegraphics{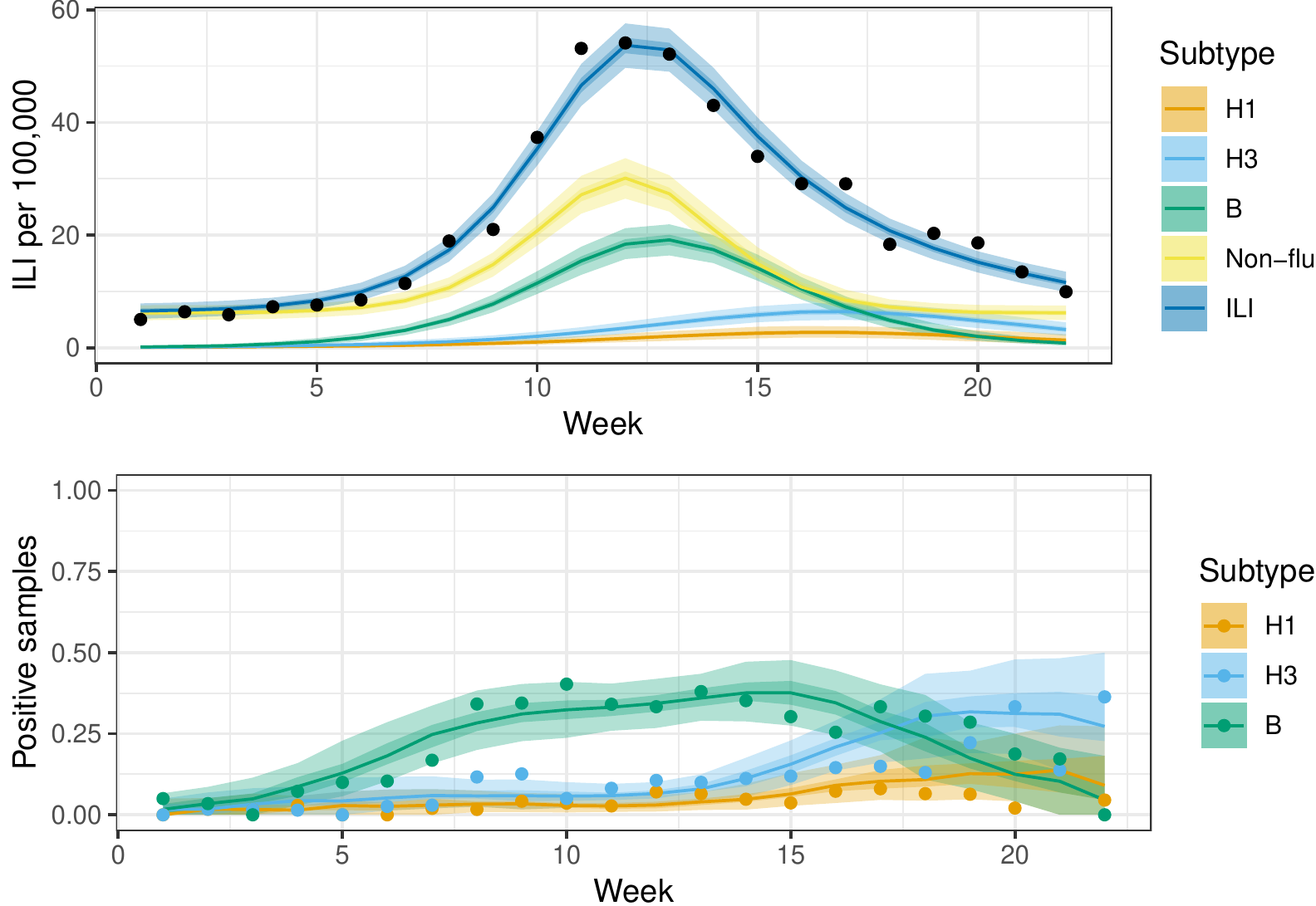}
\caption{\label{fig:plotResults}Model fit to the data. Top panel has the fit
to the ILI consultation data (blue). Furthermore, the panel highlights
the causes of ILI, i.e.~by each influenza strain or other non-flu
causes. The bottom panel has the fit to the virological confirmation
data.}
\end{figure}

For this example we used the UK influenza data from the 2017/18 season (\cite{PHE}). The 2017/18 season was somewhat unusual in that it had multiple influenza strains circulating. The main strain was a B strain, but a significant number of virological samples tested positive for the H3 strain as well. Figure \ref{fig:plotResults} shows the results of model fitting to the ILI GP consultations data and the virological confirmation data. The results show that the influenza strain causing the highest incidence is B, with also some ILI consultations due to infections with the H3 and H1 later in the season (top panel). Flu negative ILI is also an important fraction of the ILI consultations (yellow in the top panel), with a clear increase just before the B outbreak (11-13th week). For the virological confirmation the uncertainty increases after week 17, this is because later in the season less virological samples are taken, resulting in much lower confidence in the actual level of positivity by strain.

\section{Discussion}
\label{S:5}
In this paper, we summarize the basic concepts required to perform HMC and VB using Stan, in the context of infectious disease modelling. Stan is the first general purpose statistical software allowing for relatively straightforward fitting of ODE-based models using HMC and VB. In the presence of a system of ODEs, the respective likelihood function may have ridged regions resulting in a failure of standard regularity conditions and therefore difficulties in classical likelihood or MCMC-based inference. In these cases, we know that HMC may produce more accurate results and is readily available to epidemiologists in the form of Stan. 

Stan offers flexibility in the sense that it allows for the fitting to data of a very general class of models. A detailed listing of the complex models that Stan facilitates inference for is beyond the scope of this paper but can be found in the extensive documentation~\cite[]{Stan2018Stan}. In addition, one only needs to change a few lines of code in order to estimate different models either by changing the distributional assumptions or adding more components, say. Thus, as a generic and flexible software package along with the fact that it may perform inference fast, Stan makes real-time inference feasible.

We are not concerned in this article with detailed comparisons between HMC and ADVI algorithms as performed in Stan, since there are many factors that may affect their performance and certainly differ among different models. The chosen parameterization, priors, starting values and tuning parameters, are only a few of these factors. In general, HMC tends to be more computationally intensive than ADVI but it also offers high statistical efficiency. For epidemic models where the posterior distributions may be characterised by highly correlated parameter spaces, HMC seems to perform better than classical techniques.
Currently, HMC in Stan, does not allow for discrete parameters, but if they are bounded they can, in principle, be marginalized out. Finally, ADVI seems to be very promising for real-time inference but it is extremely sensitive to starting values and can underestimate posterior uncertainty. However, in practice when repeated fitting is required, say in the context of real-time inference, one may overcome this issue by a laborious initial fitting, possibly using HMC, and subsequent usage of the outcome in order to initialise the following fit.

\section*{Acknowledgements}
AC acknowledges that part of this research is co-financed by Greece and the European Union (European Social Fund- ESF) through the Operational Programme `Human Resources Development, Education and Lifelong Learning' in the context of the project `Strengthening Human Resources Research Potential via Doctorate Research' (MIS-5000432), implemented by the State Scholarships Foundation (IKY)'; ND acknowledges support from the Athens University of Economics and Business' Research Centre Action: `Original Scientific Publications'; OR from the NIH (grant number 1R01AI127232-01) and the Bill \& Melinda Gates Foundation (OPP1175094); MB thanks the MRC Centre for Global Infectious Disease Analysis (grant MR/R015600/1) and the UK National Institute for Health Research Health Protection Research Unit (NIHR HPRU) in Modelling Methodology at Imperial College London in partnership with Public Health England (PHE) (grant HPRU-2012–10080) for funding.

\appendix \label{Appendix}

\hypertarget{appA}{\section{HMC-NUTS and ADVI}}

\subsection{HMC algorithm as performed in Stan}

 \begin{itemize}
        \item {\textbf{Goal}: sample from some target density $\pi(\theta)$, where $\theta$\footnote{Note that here $\theta$ refers to the parameters of the posterior but for simplicity of notation we drop the data in this description.} is the vector of parameters of interest.}
        \item {\textbf{Auxiliary step}: 
               \begin{itemize}
                   \item {Expand the original probabilistic system by introducing auxiliary momentum parameters $p$}
                   \item {Express the target density into a joint probability distribution:\\
                   \begin{equation*}
                       \pi(p,\theta)=\pi(p|\theta)\pi(\theta)
                   \end{equation*}\\
                   which can be written in terms of the Hamiltonian as:\\
                   \begin{equation*}
                   \pi(p,\theta)=exp(-H(p,\theta))
                   \end{equation*}\\
                   thus,
                   \begin{equation*} 
                   \begin{split}
                   H(p,\theta)&=-\log{\pi}(p,\theta)\\
                              & = -\log{\pi}(p|\theta)-\log\pi(\theta)\\
                              & \equiv \underbrace{T(p,\theta)}+\underbrace{V(\theta)}\\
                   \end{split}
                \end{equation*}

                \color{white}{cccccccccccccccccccccccccc}\color{black}{$kinetic$} \color{white}{cc}\color{black}$potential$\\
                \color{white}{cccccccccccccccccccccccccc}\color{black}$energy$ \color{white}{cc}\color{black}$energy$\\
                
                and the partial derivatives of the Hamiltonian determine how $\theta$ and
                $p$ change over time, $t$, according to Hamilton's equations:\\
                \begin{equation*}
                \begin{aligned}[C]
                 &\textnormal{ }\frac{d\theta}{dt}=  \frac{\partial {H}}{\partial {p}}= \frac{\partial {T}}{\partial {p}}\\
                 &\textnormal{ }\frac{dp}{dt}=  -\frac{\partial {H}}{\partial {\theta}}= -\frac{\partial {T}}{\partial {\theta}} -\frac{\partial {V}}{\partial {\theta}}=\frac{\partial {\log\pi(p|\theta)}}{\partial {\theta}}-\frac{\partial {V}}{\partial {\theta}}=\frac{\partial {\log\pi(p)}}{\partial {\theta}}-\frac{\partial {V}}{\partial {\theta}}=-\frac{\partial {V}}{\partial {\theta}}
                \end{aligned}
                \end{equation*}
                \\
                since the density of momentum parameters is independent of the target density i.e. $\log\pi(p|\theta)=\log\pi(p)$.}
               \end{itemize}
               }
                 
    \item {\textbf{1st step}:\\
    Start from the current value of $\theta$ and draw independently a value for the momentum $p$ from a zero-mean normal distribution,\\
       \begin{equation*}
       p\sim \operatorname{MultiNormal} (0,\Sigma)
       \end{equation*}\\
       where $\Sigma$ is the covariance matrix which is also known as the mass matrix or metric \cite[]{2011arXiv1112.4118B}. The choice of $\Sigma$ can improve the efficiency of the HMC algorithm since it can rescale the target distribution so the parameters have the same scale and rotate it appropriately so the parameters are approximately independent.}
        
    \item {For \textbf{L steps} alternate half-step updates of the momentum $p$ and full-step updates of $\theta$:
        \begin{enumerate}[\indent {}]
            \item  {
             \begin{equation*}
                p \leftarrow p - \frac{\epsilon}{2} \frac{\partial {V}}{\partial {\theta}}
                \end{equation*}}
            \item{\begin{equation*}
                \theta \leftarrow \theta + \epsilon \Sigma p
                \end{equation*}}
            \item{\begin{equation*}
                p \leftarrow p - \frac{\epsilon}{2} \frac{\partial {V}}{\partial {\theta}}
                \end{equation*}}
        \end{enumerate}
    Therefore, each designed path of the algorithm has length $\epsilon L $ .The optimal choice of the step size $\epsilon$ and the number of steps $L$ play a crucial role in the performance of HMC since paths which are too short do not efficiently explore the posterior space, while paths which are too long may be rejected too often resulting in computational inefficiency. Essentially, if $\epsilon$ is too large, the leapfrog integrator’s error which depends on $\epsilon$ will be large, resulting in too many rejected proposals. If $\epsilon$ is too small then the leapfrog integrator will have to perform too many small steps, increasing run-time. On the other hand, when choosing an $L$ which is too small the proposed samples will be close to one another while choosing an $L$ which is too large, the algorithm will have to do a large number of additional computations at each iteration.}
     
    \item {\textbf{Automatic Tuning of the parameters}
    \begin{itemize}
    \item{Automatically select $L$ using the no-U-turn sampler (NUTS) in each iteration \cite[]{hoffman2014no}. NUTS uses a recursive algorithm generating an independent unit-normal
random momentum and then following a doubling procedure of leapfrog steps. Crudely, when the designed path starts to turn around, as assessed by a specific metric, NUTS stops and takes a sample. Then it generates another random momentum and initiates an additional simulation. The number of doublings is known as the tree depth and it is a control parameter~\cite[]{Stan2018Stan,2016arXiv160100225B}. So NUTS selects a sample either when the parameter space turns back on itself or when the maximum number of doublings is reached.}
    \item{Automatically determine $\epsilon$ during the warmup phase in order to match a target acceptance rate~\cite[]{Stan2018Stan,2014arXiv1411.6669B}.}
    \item{Set $\Sigma$ to be the identity matrix or restrict it to a diagonal matrix or estimate it using warmup samples~\cite[]{Stan2018Stan}.}
    \end{itemize}
    }
  \end{itemize}
  
\subsection{ADVI algorithm as performed in Stan}

\begin{itemize}
    \item {\textbf{Goal}: approximate some target density $\pi(\theta|y)$.}

    \item{{\textbf{Variational Approximation}:
    \begin{itemize}
    \item  {Consider a family of approximating densities of the latent variables \(q(\theta;\phi)\), parameterized by a vector of parameters \(\phi \in \Phi\)}
    \item  {Find the member of that family that minimizes the Kullback-Leibler(KL) divergence:
    \begin{equation*}
        \argmin_{\phi \in \Phi} KL\left(q(\theta;\phi)\|\pi(\theta|y)\right)
        \end{equation*}
       \begin{equation*}
        \textrm{such that} \quad supp(q(\theta;\phi))\subseteq supp(\pi(\theta|y))
        \end{equation*}
        where $y$ denotes the data.
        
        }
    \item {Since,
    \begin{equation*}
    \begin{split}
     KL\left(q(\theta;\phi)\|\pi(\theta|y)\right)& =\mathbb{E}_{q(\theta)}[\log q(\theta;\phi)]-\mathbb{E}_{q(\theta)}[\log \pi(\theta|y)]\\
            & =\mathbb{E}_{q(\theta)}[\log q(\theta;\phi)]-\mathbb{E}_{q(\theta)}[\log \pi(y,\theta)]+\mathbb{E}_{q(\theta)}[\log\pi(y)]\\
            & =-\underbrace{\left[\mathbb{E}_{q(\theta)}[\log \pi(y,\theta)]-\mathbb{E}_{q(\theta)}[\log q(\theta;\phi)]\right]}+\log\pi(y)
    \end{split}
    \end{equation*}
     \color{white}{ccccccccccccccccccccccccccccccccccccccc}\color{black}{ELBO}\\
    
    so the KL divergence involves the target density and its analytic form is unknown. However, notice that $\log\pi(y)$ does not depend on the variational density $q(\theta)$, so it is a constant. Thus, minimizing the KL divergence is equivalent to minimizing the Evidence Lower Bound (ELBO):
    \begin{equation*}
    \argmax_{\phi \in \Phi}\left[\mathbb{E}_{q(\theta)}[\log \pi(y,\theta)]-\mathbb{E}_{q(\theta)}[\log q(\theta;\phi)]\right]
    \end{equation*}
    subject to the support constraint.}    
    \end{itemize}
    }}
\item{\textbf{1st step}:
Transform the parameters of interest, $T:\theta\rightarrow\zeta$, so that their support is in the real coordinate space i.e. define a one-to-one differentiable function, \( T: supp(\pi(\theta))\rightarrow \mathbb{R}^{\kappa}\).
Then the transformed density is denoted by:
\begin{equation*} 
\begin{split}
\pi(y,\zeta)& =\pi\left(y,T^{-1}(\zeta)\right)|\operatorname{det} J_{T^{-1}}(\zeta)|\\
            & =\pi(y,\theta)|\operatorname{det} J_{T^{-1}}(\zeta)|
\end{split}
\end{equation*}
where $J_{T^{-1}}(\zeta)$ is the Jacobian of the inverse of $T$.\\
Stan supports and automatically uses a library of transformations and their corresponding Jacobians.\\
Also, it can be shown that the ELBO in the real coordinate space is:
\begin{equation*}
    \mathcal{L(\phi)}=\mathbb{E}_{q(\zeta;\phi)}\left[\log \pi\left(y,T^{-1}(\zeta)\right)+\log|\operatorname{det} J_{T^{-1}}(\zeta)|]-\mathbb{E}_{q(\zeta;\phi)}[\log q(\zeta;\phi)\right]
    \end{equation*}
}
\item {\textbf{2nd step}: Choose the variational approximation
\begin{itemize}
    \item {Mean-field or factorized Gaussian
    \begin{equation*}
        q(\zeta;\phi)=\displaystyle \prod_{\kappa=1}^{K} \mathcal{N}(\zeta_{\kappa};\mu_{\kappa},\sigma_{\kappa}^2)
    \end{equation*}
    where $\phi=(\mu_{1},\dots,\mu_{K},\sigma_{1}^2,\dots,\sigma_{K}^2)$.
    }
    \item {Full-rank Gaussian
    \begin{equation*}
        q(\zeta;\phi)= \mathcal{N}(\zeta;\mu,\Sigma)
    \end{equation*}}
    where $\phi=(\mu,\Sigma)$.
\end{itemize}
}
\item {\textbf{3rd step}: Stochastic optimization in order to maximize the ELBO in the real coordinate space~\cite[]{kucukelbir2017automatic}:
\begin{itemize}
    \item {The expectations with respect to the variational parameters $\phi$ constituting the ELBO, are unknown. Apply an elliptical standardization so the expectations do not depend on $\phi$.}
    \item {Compute the gradients inside the expectation with automatic differentiation and use Monte Carlo integration to compute the expectations.}
    \item {Given the gradients of the ELBO employ a stochastic gradient ascent algorithm.}
\end{itemize}}

\end{itemize}

\hypertarget{appCοde}{\section{Stan model code and implementation}}
A Stan model consists of a number of “blocks”,  where variables are declared by their type according to their use. All variables should have a declared data type and size. This should be done at the start of each block. Also, local variables can be declared at the beginning of each block. The primitive types represent real and integer values while vectors, row vectors, and matrices as well as arrays are also supported. Vector and matrix types necessarily contain only real values, so collections of integers are expressed using arrays. The declared variables can be constrained given lower and upper bounds which should be imposed carefully.

A complete Stan model is composed of six code blocks named \texttt{data}, \texttt{transformed data}, \texttt{parameters}, \texttt{transformed parameters} and \texttt{generated quantities}. There is also a \texttt{functions}-definition block where user-defined functions are constructed and if used, this block should appear before all of the other program blocks. In general, the declarations and statements which constitute the Stan program, are executed in the order in which they are written so everything should be stated consistently. The \texttt{data} block consists of the data required to fit the model while the \texttt{transformed data} block may include temporary transformations of the data, independent of the parameters, which need to be saved. The model's parameters which the user want to infer are defined in the \texttt{parameters} and in the \texttt{transformed parameters} blocks. Intermediate variables can be declared in terms of data and parameters. These values will also be returned by the inference based on the draws from the posterior parameters. The \texttt{model} block is the core of Stan model statement and is where the model is defined in terms of priors and likelihood. Sampling statements can be used but log probability variables can also be accessed directly, or user-defined probability functions can be employed. Finally, the \texttt{generated quantities} block may be used to define quantities that depend on parameters and data or even random number generation and don't affect inference.  

In what follows we illustrate a complete Stan model. However, the reader is referred to \url{https://mc-stan.org/} for the latest official Stan documentation for detailed instructions. Code for all the examples employed in this paper is made freely available in  \url{https://github.com/anastasiachtz/COMMAND_stan.git}. Here, we demonstrate Stan model code by fitting the single strain deterministic model to data for an influenza outbreak in a boarding school in the north of England. The model as described by equations (\ref{eq:ref5})-(\ref{eq:ref6}) can be written in Stan in the following form, which the user should save as .stan file:
\begin{lstlisting}[language=HTML]
functions {
  real[] SIR(real t,  // time
  real[] y,           // system state {susceptible,infected,recovered}
  real[] theta,       // parameters {transmission rate, recovery rate}
  real[] x_r,         // real valued fixed data
  int[] x_i) {        // integer valued fixed data

  real dy_dt[3];

  dy_dt[1] = - theta[1] * y[1] * y[2];
  dy_dt[2] = theta[1] * y[1] * y[2] - theta[2] * y[2];
  dy_dt[3] = theta[2] * y[2];

  return dy_dt;
}

}
data {
  int<lower = 1> n_obs;    // number of days observed
  int<lower = 1> n_theta;  // number of model parameters
  int<lower = 1> n_difeq;  // number of differential equations
  int<lower = 1> n_pop;    // population 
  int y[n_obs];            // data, total number of infected each day
  real t0;                 // initial time point (zero)
  real ts[n_obs];          // time points observed
}

transformed data {
  real x_r[0];
  int x_i[0];
}

parameters {
  real<lower = 0> theta[n_theta];   // model parameters 
  real<lower = 0, upper = 1> S0;    // initial fraction of susceptible
}

transformed parameters{
  real y_hat[n_obs, n_difeq]; // solution from the ODE solver
  real y_init[n_difeq];       // initial conditions for both susceptible
                              // and infected

  y_init[1] = S0;
  y_init[2] = 1 - S0;
  y_init[3] = 0;

  y_hat = integrate_ode_rk45(SIR, y_init, t0, ts, theta, x_r, x_i);
}

model {
  real lambda[n_obs];    // Poisson parameter

  //priors
  theta[1] ~ lognormal(0,1);
  theta[2] ~ gamma(0.004,0.02); 
  S0 ~ beta(0.5, 0.5);

  //likelihood
  for (i in 1:n_obs){
   lambda[i] = y_hat[i,2]*n_pop;
  }
  y ~ poisson(lambda);
}

generated quantities {
  real R_0;             // Basic reproduction number
  R_0 = theta[1]/theta[2];
}

\end{lstlisting}
In the \texttt{functions} block, the system of ODEs is coded directly in Stan as a function with a strictly specified signature. It takes as input time, system state, parameters and real and integer data, in exactly this order, and returns the derivatives with respect to time. Note that, the initial state can also be estimated along with the parameters describing the system, which is also done here. In order to solve the system Stan has two built-in ODE solvers, \texttt{integrate\_ode\_rk45} and \texttt{integrate\_ode\_bdf}. Both take similar variables and functions, but they take solver specific arguments as well. The first argument must be the function that describes the ODE system but the other arguments, except for the initial state and the parameters, are restricted to data only expressions already declared. The solutions to the ODEs describing the SIR, given initial conditions, are defined in the block of \texttt{transformed parameters}. These intermediate values can be used in the model section and the posterior values will be included in the stan output.

Once the .stan file is written, the user should load the necessary libraries, provide data and fit the model. To do so, we use the R interface to Stan. For this implementation we use data from the R package \texttt{outbreaks}, maintained as part of the R Epidemics Consortium (RECON; \url{http://www.repidemicsconsortium.org}).

\begin{lstlisting}
library(deSolve)
library(dplyr)
library(rstan)
library(outbreaks)

# Automatically save compiled Stan models so they can be ran multiple 
# times without getting recompiled:
rstan_options(auto_write = TRUE)

# Chains will run in parallel when possible:
options(mc.cores = parallel::detectCores())

onset <- influenza_england_1978_school$date    # Onset date
cases <- influenza_england_1978_school$in_bed  # Number of infected students
N = length(onset)   # Number of days observed throughout the outbreak
pop = 763           # Population 
sample_time=1:N

# Modify data into a form suitable for Stan
flu_data = list(n_obs = N,
                n_theta = 2,
                n_difeq = 3,
                n_pop = pop,
                y = cases,
                t0 = 0,
                ts = sample_time)

# Specify parameters to monitor
parameters = c("y_hat", "y_init", "theta",  "R_0")
\end{lstlisting}
\vspace{.7cm}
Fit the model using the default algorithm, NUTS:
\begin{lstlisting}
n_chains=5
n_warmups=500
n_iter=100500
n_thin=50
set.seed(1234)

# Set initial values:
ini = function(){
  list(theta=c(runif(1,0,5), runif(1,0.2,0.4)), 
       S0=runif(1,(pop-3)/pop,(pop-1)/pop))  
}

nuts_fit = stan(file = "SIR_det_Poisson.stan", # Stan program
                data = flu_data,     # list of data
                pars = parameters,   # monitored parameters
                init = ini,          # initial parameter values
                chains = n_chains,   # number of Markov chains to run
                warmup = n_warmups,  # number of warmup iterations per chain
                iter = n_iter,    # number of iterations per chain(+warmup)
                thin=n_thin,      # period for saving samples
                seed=13219)
\end{lstlisting}
By default, Stan generates its own initial values randomly between -2 and 2 for each parameter. However, especially in complex models as those including non-linear systems of ODEs, it is better to specify the initial values for at least a subset of the parameters. Except for initial values, the length of adaptation during the warm-up phase is also important since at this step Stan tries to find the appropriate step size of the leapfrog integrator which will result in efficient sampling and at the same time avoid failures of the integrator, identified as divergences. The step size is determined trying to achieve a target acceptance rate which is specified by a \texttt{adapt\_delta} argument in the \texttt{stan()} function which is also a tuning parameter for the algorithm. In this example, the default value of 0.8 is used for \texttt{adapt\_delta}. In general, Stan indicates if there are divergences so the user can increase the value of \texttt{adapt\_delta} getting closer to its maximum value of 1, decreasing in this way the step size if needed.

The \texttt{stan()} function returns a stanfit object which contains the sample drawn from the posterior for the monitored parameters. Printing the stanfit object will automatically evaluate the estimated mean, standard error of the mean, standard deviation, percentiles, effective sample size and $\hat{R}$ statistic for each parameter. The stanfit object can also interface with some R commands like summary so we can inspect specific parameters of interest.
\begin{lstlisting}
print(nuts_fit)

nuts_fit_summary <- summary(nuts_fit, pars = c("lp_", "theta[1]","theta[2]",
                            "y_init[1]", "R_0"))$summary
print(nuts_fit_summary,scientific=FALSE,digits=2)

# Obtain the generated samples:
posts <-  rstan::extract(nuts_fit)
\end{lstlisting}
\begin{figure}[H]
  \centering
  \includegraphics[width=1\linewidth]{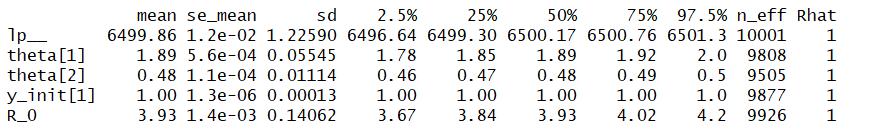}
  \caption*{\label{fig:nuts_summ}}
\end{figure}
\vspace{-1.1cm}
Additional diagnostics such as checking for divergent transitions and inspecting the maximum trajectory length are also available. As mentioned earlier, failures of the leapfrog integrator are identified as divergences. In cases where the parameter space is not well behaved, NUTS may move according to the dynamically selected step size until it hits the maximum number of leapfrog doublings, known as tree depth. However, this means that the algorithm will select draws according to this threshold, instead of actually tracing the posterior, so the user should check whether there are iterations where the treedepth exceeds the maximum. Note that, problematic specification of the model may always be the source of divergences and reparameterizations should be considered.
\begin{lstlisting}
# Inspect all the values of parameters used for the sampler per chain:
sampler_params <- get_sampler_params(fit, inc_warmup = FALSE)

check_divergences(nuts_fit)
# 0 of 10000 iterations ended with a divergence.

check_treedepth(nuts_fit)
# 0 of 10000 iterations saturated the maximum tree depth of 10.
\end{lstlisting}
Using the bayesplot package the user can obtain trace plots of the fit, to assess the convergence of chains, univariate and bivariate marginal posterior distributions as well as other diagnostics (see~\cite{gabry2019visualization}). The user should always examine model diagnostics in more detail especially in more complex models, here we illustrate just some preliminary steps.
\begin{lstlisting}
library(bayesplot)

posterior <- as.array(nuts_fit)

mcmc_trace(posterior_1, pars=c("lp_", "theta[1]", "theta[2]", 
                               "y_init[1]", "R_0"))

pairs(nuts_fit_2, pars = c("theta[1]", "theta[2]", "y_init[1]"), 
                            labels = c("beta", "gamma", "s(0)"), 
                            cex.labels=1.5, font.labels=9, 
                            condition = "accept_stat__")                    
\end{lstlisting}

\vspace{.3cm}
Given the already specified stan model, we can fit the model using ADVI simply by calling the function \texttt{vb()}. In this example we use the default setting which performs mean-field ADVI, using the credible intervals we obtained from NUTS as initial values:
\begin{lstlisting}
# Set initial values:
ini_vb = function(){
  list(params=c(runif(1,1.85,1.92), runif(1,0.47,0.49)), 
       S0=runif(1,(pop-2)/pop,(pop-1)/pop))}

mod=stan_model("SIR_det_Poisson.stan")

fit_vb=vb(mod, 
          data = flu_data, 
          pars = parameters, 
          init = ini_vb, 
          iter = 10000, 
          tol_rel_obj = 0.001,
          seed=16735679)
\end{lstlisting}
Stan reports the average and median changes of the ELBO during the stochastic optimization and if either don’t fall below a certain threshold of \texttt{tol\_rel\_obj} then the algorithm has converged. Currently, we can't actually check the performance of ADVI, however there is ongoing research on diagnostics for variational inference algorithms~\cite[]{yao2018yes}.

The \texttt{vb()} function returns a stanfit object which contains the approximate draws from the posterior for the monitored parameters and printing it automatically evaluates the approximated mean, standard deviation and percentiles. 
\begin{lstlisting}
print(vb_fit)

vb_fit_summary <- summary(vb_fit, pars = c("theta[1]", "theta[2]",
                          "y_init[1]", "R_0"))$summary
print(vb_fit_summary,scientific=FALSE,digits=2)

# Extract the approximate samples:
posts_vb <-  rstan::extract(vb_fit)
\end{lstlisting}
\begin{figure}[H]
  \centering
  \includegraphics[width=.7\linewidth]{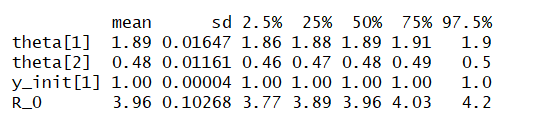}
  \caption*{\label{fig:advi_summ}}
\end{figure}

A basic advantage of Stan is flexibility in modeling, as we only need to change a few lines of code in order to implement different models, either by changing the distributional assumptions or adding more components. For example, in the setting of the single strain deterministic SIR, we can also use a Binomial likelihood simply by changing one line of code. For example we consider a Binomial model, using the same prior distributions, formulated as follows:\\
\begin{equation}
    \textnormal{ $Y_{t}\sim \operatorname{Bin} \left({N, p_t}\right)$}\\
\end{equation}
\begin{equation}
    \textnormal{$p_t=\int_{0}^{t}\left({\beta i_s s_s - \gamma i_s}\right) \mathrm{d}s$}
\end{equation}
where $s_s$ is the fraction of susceptible students and $i_s$ is the fraction of infected students.\\
In order to write the model in Stan we need to change only the model block:
\begin{lstlisting}[language=HTML]
model {
  //priors
  theta[1] ~ lognormal(0,1);
  theta[2] ~ gamma(0.004,0.02); 
  S0 ~ beta(0.5, 0.5);
  
  //likelihood
  y ~ binomial(n_pop, y_hat[, 2]);
 }
\end{lstlisting}
We would save the new .stan file and perform inference using NUTS and ADVI as before.

\newpage
\hypertarget{references}{%
\label{references}}    

\bibliography{references.bib}

\begin{thebibliography}{51}
\providecommand{\natexlab}[1]{#1}
\providecommand{\url}[1]{\texttt{#1}}
\expandafter\ifx\csname urlstyle\endcsname\relax
  \providecommand{\doi}[1]{doi: #1}\else
  \providecommand{\doi}{doi: \begingroup \urlstyle{rm}\Url}\fi

\bibitem[Anderson and May(1992)]{anderson1992infectious}
R.~M. Anderson and R.~M. May.
\newblock \emph{Infectious diseases of humans: dynamics and control}.
\newblock Oxford university press, 1992.

\bibitem[Andersson and Britton(2000)]{andersson2012stochastic}
H.~Andersson and T.~Britton.
\newblock \emph{Stochastic epidemic models and their statistical analysis},
  volume 151.
\newblock Springer Science \& Business Media, 2000.

\bibitem[Baguelin et~al.(2013)Baguelin, Flasche, Camacho, Demiris, Miller, and
  Edmunds]{baguelin2013assessing}
M.~Baguelin, S.~Flasche, A.~Camacho, N.~Demiris, E.~Miller, and W.~J. Edmunds.
\newblock Assessing optimal target populations for influenza vaccination
  programmes: an evidence synthesis and modelling study.
\newblock \emph{PLoS medicine}, 10\penalty0 (10):\penalty0 e1001527, 2013.
\newblock \doi{10.1371/journal.pmed.1001527}.

\bibitem[Beskos et~al.(2013)Beskos, Pillai, Roberts, Sanz-Serna, Stuart,
  et~al.]{beskos2013optimal}
A.~Beskos, N.~Pillai, G.~Roberts, J.-M. Sanz-Serna, A.~Stuart, et~al.
\newblock {Optimal tuning of the hybrid Monte Carlo algorithm}.
\newblock \emph{Bernoulli}, 19\penalty0 (5A):\penalty0 1501--1534, 2013.

\bibitem[{Betancourt}(2016)]{2016arXiv160100225B}
M.~{Betancourt}.
\newblock {Identifying the Optimal Integration Time in Hamiltonian Monte
  Carlo}.
\newblock \emph{arXiv e-prints}, art. arXiv:1601.00225, Jan 2016.

\bibitem[{Betancourt}(2017)]{2017arXiv170102434B}
M.~{Betancourt}.
\newblock {A Conceptual Introduction to Hamiltonian Monte Carlo}.
\newblock \emph{arXiv e-prints}, art. arXiv:1701.02434, Jan 2017.

\bibitem[{Betancourt} and {Stein}(2011)]{2011arXiv1112.4118B}
M.~{Betancourt} and L.~C. {Stein}.
\newblock {The Geometry of Hamiltonian Monte Carlo}.
\newblock \emph{arXiv e-prints}, art. arXiv:1112.4118, Dec 2011.

\bibitem[Betancourt et~al.(2017)Betancourt, Byrne, Livingstone, Girolami,
  et~al.]{betancourt2017geometric}
M.~Betancourt, S.~Byrne, S.~Livingstone, M.~Girolami, et~al.
\newblock {The geometric foundations of Hamiltonian Monte Carlo}.
\newblock \emph{Bernoulli}, 23\penalty0 (4A):\penalty0 2257--2298, 2017.
\newblock \doi{10.3150/16-BEJ810}.

\bibitem[{Betancourt} et~al.(2014){Betancourt}, {Byrne}, and
  {Girolami}]{2014arXiv1411.6669B}
M.~J. {Betancourt}, S.~{Byrne}, and M.~{Girolami}.
\newblock {Optimizing The Integrator Step Size for Hamiltonian Monte Carlo}.
\newblock \emph{arXiv e-prints}, art. arXiv:1411.6669, Nov 2014.

\bibitem[Bishop(2006)]{bishop2006pattern}
C.~M. Bishop.
\newblock \emph{Pattern recognition and machine learning}.
\newblock springer, 2006.

\bibitem[Blei et~al.(2017)Blei, Kucukelbir, and McAuliffe]{blei2017variational}
D.~M. Blei, A.~Kucukelbir, and J.~D. McAuliffe.
\newblock {Variational inference: A review for statisticians}.
\newblock \emph{Journal of the American Statistical Association}, 112\penalty0
  (518):\penalty0 859--877, 2017.
\newblock \doi{10.1080/01621459.2017.1285773}.

\bibitem[{Carpenter} et~al.(2015){Carpenter}, {Hoffman}, {Brubaker}, {Lee},
  {Li}, and {Betancourt}]{2015arXiv150907164C}
B.~{Carpenter}, M.~D. {Hoffman}, M.~{Brubaker}, D.~{Lee}, P.~{Li}, and
  M.~{Betancourt}.
\newblock {The Stan Math Library: Reverse-Mode Automatic Differentiation in
  C++}.
\newblock \emph{arXiv e-prints}, art. arXiv:1509.07164, Sep 2015.

\bibitem[Carpenter et~al.(2017)Carpenter, Gelman, Hoffman, Lee, Goodrich,
  Betancourt, Brubaker, Guo, Li, and Riddell]{carpenter2017Stan}
B.~Carpenter, A.~Gelman, M.~D. Hoffman, D.~Lee, B.~Goodrich, M.~Betancourt,
  M.~Brubaker, J.~Guo, P.~Li, and A.~Riddell.
\newblock {Stan: A probabilistic programming language}.
\newblock \emph{Journal of statistical software}, 76\penalty0 (1), 2017.
\newblock \doi{10.18637/jss.v076.i01}.

\bibitem[de~Valpine et~al.(2017)de~Valpine, Turek, Paciorek, Anderson-Bergman,
  Lang, and Bodik]{de2017programming}
P.~de~Valpine, D.~Turek, C.~J. Paciorek, C.~Anderson-Bergman, D.~T. Lang, and
  R.~Bodik.
\newblock {Programming with models: writing statistical algorithms for general
  model structures with NIMBLE}.
\newblock \emph{Journal of Computational and Graphical Statistics}, 26\penalty0
  (2):\penalty0 403--413, 2017.
\newblock \doi{10.1080/10618600.2016.1172487}.

\bibitem[De~Vries et~al.(2006)De~Vries, Hillen, Lewis, Sch{\'O}nfisch,
  et~al.]{de2006course}
G.~De~Vries, T.~Hillen, M.~Lewis, B.~Sch{\'O}nfisch, et~al.
\newblock \emph{A course in mathematical biology: quantitative modeling with
  mathematical and computational methods}, volume~12.
\newblock Siam, 2006.

\bibitem[Fleming and Elliot(2008)]{fleming2008lessons}
D.~Fleming and A.~Elliot.
\newblock {Lessons from 40 years' surveillance of influenza in England and
  Wales}.
\newblock \emph{Epidemiology \& Infection}, 136\penalty0 (7):\penalty0
  866--875, 2008.
\newblock \doi{10.1017/S0950268807009910}.

\bibitem[Fournier et~al.(2012)Fournier, Skaug, Ancheta, Ianelli, Magnusson,
  Maunder, Nielsen, and Sibert]{fournier2012ad}
D.~A. Fournier, H.~J. Skaug, J.~Ancheta, J.~Ianelli, A.~Magnusson, M.~N.
  Maunder, A.~Nielsen, and J.~Sibert.
\newblock {AD Model Builder: using automatic differentiation for statistical
  inference of highly parameterized complex nonlinear models}.
\newblock \emph{Optimization Methods and Software}, 27\penalty0 (2):\penalty0
  233--249, 2012.

\bibitem[Gabry et~al.(2019)Gabry, Simpson, Vehtari, Betancourt, and
  Gelman]{gabry2019visualization}
J.~Gabry, D.~Simpson, A.~Vehtari, M.~Betancourt, and A.~Gelman.
\newblock {Visualization in Bayesian workflow}.
\newblock \emph{Journal of the Royal Statistical Society: Series A (Statistics
  in Society)}, 182\penalty0 (2):\penalty0 389--402, 2019.

\bibitem[Gelman and Hill(2006)]{gelman2006data}
A.~Gelman and J.~Hill.
\newblock \emph{Data analysis using regression and multilevel/hierarchical
  models}.
\newblock Cambridge university press, 2006.

\bibitem[Gelman et~al.(2013)Gelman, Stern, Carlin, Dunson, Vehtari, and
  Rubin]{gelman2013bayesian}
A.~Gelman, H.~S. Stern, J.~B. Carlin, D.~B. Dunson, A.~Vehtari, and D.~B.
  Rubin.
\newblock \emph{Bayesian data analysis}.
\newblock Chapman and Hall/CRC, 2013.

\bibitem[Geman and Geman(1984)]{geman1984stochastic}
S.~Geman and D.~Geman.
\newblock {Stochastic relaxation, Gibbs distributions, and the Bayesian
  restoration of images}.
\newblock \emph{IEEE Transactions on pattern analysis and machine
  intelligence}, \penalty0 (6):\penalty0 721--741, 1984.
\newblock \doi{10.1109/TPAMI.1984.4767596}.

\bibitem[Griewank and Walther(2008)]{griewank2008evaluating}
A.~Griewank and A.~Walther.
\newblock \emph{Evaluating derivatives: principles and techniques of
  algorithmic differentiation}, volume 105.
\newblock Siam, 2008.

\bibitem[Griewank et~al.(1989)]{griewank1989automatic}
A.~Griewank et~al.
\newblock On automatic differentiation.
\newblock \emph{Mathematical Programming: recent developments and
  applications}, 6\penalty0 (6):\penalty0 83--107, 1989.

\bibitem[Hastings(1970)]{hastings1970monte}
W.~K. Hastings.
\newblock {Monte Carlo sampling methods using Markov chains and their
  applications}.
\newblock 1970.
\newblock \doi{10.1093/biomet/57.1.97}.

\bibitem[Hoffman and Gelman(2014)]{hoffman2014no}
M.~D. Hoffman and A.~Gelman.
\newblock {The No-U-Turn sampler: adaptively setting path lengths in
  Hamiltonian Monte Carlo.}
\newblock \emph{Journal of Machine Learning Research}, 15\penalty0
  (1):\penalty0 1593--1623, 2014.

\bibitem[Jordan et~al.(1999)Jordan, Ghahramani, Jaakkola, and
  Saul]{jordan1999introduction}
M.~I. Jordan, Z.~Ghahramani, T.~S. Jaakkola, and L.~K. Saul.
\newblock An introduction to variational methods for graphical models.
\newblock \emph{Machine learning}, 37\penalty0 (2):\penalty0 183--233, 1999.
\newblock \doi{10.1023/A:1007665907178}.

\bibitem[Karatzas and Shreve(1998)]{karatzas1998brownian}
I.~Karatzas and S.~E. Shreve.
\newblock Brownian motion.
\newblock In \emph{Brownian Motion and Stochastic Calculus}, pages 47--127.
  Springer, 1998.

\bibitem[Kermack and McKendrick(1927)]{kermack1927contribution}
W.~O. Kermack and A.~G. McKendrick.
\newblock A contribution to the mathematical theory of epidemics.
\newblock \emph{Proceedings of the royal society of london. Series A,
  Containing papers of a mathematical and physical character}, 115\penalty0
  (772):\penalty0 700--721, 1927.
\newblock \doi{10.1098/rspa.1927.0118}.

\bibitem[Kristensen et~al.(2015)Kristensen, Nielsen, Berg, Skaug, and
  Bell]{kristensen2015tmb}
K.~Kristensen, A.~Nielsen, C.~W. Berg, H.~Skaug, and B.~Bell.
\newblock {TMB: automatic differentiation and Laplace approximation}.
\newblock \emph{arXiv preprint arXiv:1509.00660}, 2015.

\bibitem[{Kucukelbir} et~al.(2015){Kucukelbir}, {Ranganath}, {Gelman}, and
  {Blei}]{2015arXiv150603431K}
A.~{Kucukelbir}, R.~{Ranganath}, A.~{Gelman}, and D.~M. {Blei}.
\newblock {Automatic Variational Inference in Stan}.
\newblock \emph{arXiv e-prints}, art. arXiv:1506.03431, Jun 2015.

\bibitem[Kucukelbir et~al.(2017)Kucukelbir, Tran, Ranganath, Gelman, and
  Blei]{kucukelbir2017automatic}
A.~Kucukelbir, D.~Tran, R.~Ranganath, A.~Gelman, and D.~M. Blei.
\newblock Automatic differentiation variational inference.
\newblock \emph{The Journal of Machine Learning Research}, 18\penalty0
  (1):\penalty0 430--474, 2017.

\bibitem[Kullback(1997)]{kullback1997information}
S.~Kullback.
\newblock \emph{Information theory and statistics}.
\newblock Courier Corporation, 1997.

\bibitem[Lunn et~al.(2012)Lunn, Jackson, Best, Spiegelhalter, and
  Thomas]{lunn2012bugs}
D.~Lunn, C.~Jackson, N.~Best, D.~Spiegelhalter, and A.~Thomas.
\newblock \emph{{The BUGS book: A practical introduction to Bayesian
  analysis}}.
\newblock Chapman and Hall/CRC, 2012.

\bibitem[Lunn et~al.(2000)Lunn, Thomas, Best, and
  Spiegelhalter]{lunn2000winbugs}
D.~J. Lunn, A.~Thomas, N.~Best, and D.~Spiegelhalter.
\newblock {WinBUGS-a Bayesian modelling framework: concepts, structure, and
  extensibility}.
\newblock \emph{Statistics and computing}, 10\penalty0 (4):\penalty0 325--337,
  2000.

\bibitem[Malesios et~al.(2017)Malesios, Demiris, Kalogeropoulos, and
  Ntzoufras]{malesios2017bayesian}
C.~Malesios, N.~Demiris, K.~Kalogeropoulos, and I.~Ntzoufras.
\newblock Bayesian epidemic models for spatially aggregated count data.
\newblock \emph{Statistics in medicine}, 36\penalty0 (20):\penalty0 3216--3230,
  2017.
\newblock \doi{10.1002/sim.7364}.

\bibitem[McElreath(2012)]{mcelreath2012rethinking}
R.~McElreath.
\newblock {rethinking: Statistical Rethinking book package}.
\newblock \emph{R package version}, 1, 2012.

\bibitem[Metropolis et~al.(1953)Metropolis, Rosenbluth, Rosenbluth, Teller, and
  Teller]{metropolis1953equation}
N.~Metropolis, A.~W. Rosenbluth, M.~N. Rosenbluth, A.~H. Teller, and E.~Teller.
\newblock Equation of state calculations by fast computing machines.
\newblock \emph{The journal of chemical physics}, 21\penalty0 (6):\penalty0
  1087--1092, 1953.
\newblock \doi{10.1063/1.1699114}.

\bibitem[Monnahan et~al.(2017)Monnahan, Thorson, and
  Branch]{monnahan2017faster}
C.~C. Monnahan, J.~T. Thorson, and T.~A. Branch.
\newblock {Faster estimation of Bayesian models in ecology using Hamiltonian
  Monte Carlo}.
\newblock \emph{Methods in Ecology and Evolution}, 8\penalty0 (3):\penalty0
  339--348, 2017.

\bibitem[Neal(1993)]{neal1993probabilistic}
R.~M. Neal.
\newblock {Probabilistic inference using Markov chain Monte Carlo methods}.
\newblock 1993.

\bibitem[{Neal}(2012)]{2012arXiv1206.1901N}
R.~M. {Neal}.
\newblock {MCMC using Hamiltonian dynamics}.
\newblock \emph{arXiv e-prints}, art. arXiv:1206.1901, Jun 2012.

\bibitem[O’Neill and Roberts(1999)]{o1999bayesian}
P.~D. O’Neill and G.~O. Roberts.
\newblock Bayesian inference for partially observed stochastic epidemics.
\newblock \emph{Journal of the Royal Statistical Society: Series A (Statistics
  in Society)}, 162\penalty0 (1):\penalty0 121--129, 1999.
\newblock \doi{10.1111/1467-985X.00125/}.

\bibitem[Patil et~al.(2010)Patil, Huard, and Fonnesbeck]{patil2010pymc}
A.~Patil, D.~Huard, and C.~J. Fonnesbeck.
\newblock {PyMC: Bayesian stochastic modelling in Python}.
\newblock \emph{Journal of statistical software}, 35\penalty0 (4):\penalty0 1,
  2010.

\bibitem[Plummer(2017)]{jags2017}
M.~Plummer.
\newblock {JAGS Version 4.3.0 user manual}.
\newblock \url{http://mcmc-jags.sourceforge.net/}, 2017.

\bibitem[Plummer et~al.(2003)]{plummer2003jags}
M.~Plummer et~al.
\newblock {JAGS: A program for analysis of Bayesian graphical models using
  Gibbs sampling}.
\newblock In \emph{Proceedings of the 3rd international workshop on distributed
  statistical computing}, volume 124. Vienna, Austria, 2003.

\bibitem[{Public Health England}()]{PHE}
{Public Health England}.
\newblock {Surveillance of influenza and other respiratory viruses in the {UK}:
  Winter 2017 to 2018, {PHE} publications gateway number: 2018093}.

\bibitem[{Stan Development Team}(2018)]{Stan2018Stan}
{Stan Development Team}.
\newblock {Stan Modeling Language User's Guide and Reference Manual, Version
  2.18.0}.
\newblock \url{http://mc-stan.org/}, 2018.

\bibitem[Turner and Sahani(2011)]{turner2011two}
R.~E. Turner and M.~Sahani.
\newblock Two problems with variational expectation maximization for
  time-series models.
\newblock \emph{Bayesian Time series models}, 1\penalty0 (3.1):\penalty0 3--1,
  2011.
\newblock \doi{10.1017/CBO9780511984679.006}.

\bibitem[Wainwright et~al.(2008)Wainwright, Jordan,
  et~al.]{wainwright2008graphical}
M.~J. Wainwright, M.~I. Jordan, et~al.
\newblock Graphical models, exponential families, and variational inference.
\newblock \emph{Foundations and Trends{\textregistered} in Machine Learning},
  1\penalty0 (1--2):\penalty0 1--305, 2008.
\newblock \doi{10.1561/2200000001}.

\bibitem[Wang and Blei(2018)]{wang2018frequentist}
Y.~Wang and D.~M. Blei.
\newblock {Frequentist Consistency of Variational Bayes}.
\newblock \emph{Journal of the American Statistical Association}, \penalty0
  (just-accepted):\penalty0 1--85, 2018.
\newblock \doi{10.1080/01621459.2018.1473776}.

\bibitem[Wearing et~al.(2005)Wearing, Rohani, and
  Keeling]{wearing2005appropriate}
H.~J. Wearing, P.~Rohani, and M.~J. Keeling.
\newblock Appropriate models for the management of infectious diseases.
\newblock \emph{PLoS medicine}, 2\penalty0 (7):\penalty0 e174, 2005.
\newblock \doi{10.1371/journal.pmed.0020174}.

\bibitem[Yao et~al.(2018)Yao, Vehtari, Simpson, and Gelman]{yao2018yes}
Y.~Yao, A.~Vehtari, D.~Simpson, and A.~Gelman.
\newblock {Yes, but did it work?: Evaluating variational inference}.
\newblock \emph{arXiv preprint arXiv:1802.02538}, 2018.

\end{thebibliography}

\end{document}